\documentclass[authoryear,3p]{elsarticle}

\journal{ISA Transactions}

\usepackage{graphicx}

\usepackage{amssymb,amsmath}
\usepackage{enumitem}
\usepackage{mathtools}
\usepackage{multirow}
\usepackage{subcaption, tabularx}
\usepackage{comment}
\usepackage[graphicx]{realboxes}
\usepackage{url}
\usepackage[hyperindex,pdftex]{hyperref}

\usepackage{xcolor}

\usepackage{comment}
\usepackage{algorithm,algorithmic}
\usepackage{placeins}

\newdefinition{definition}{Definition}
\newdefinition{assumption}{Assumption}
\newdefinition{remark}{Remark}

\begin{document}

\begin{frontmatter}




\title{A Parametrized Nonlinear Predictive Control Strategy for \\Relaxing COVID-19 Social
  Distancing Measures in Brazil}

\author[adUFSC]{Marcelo M. Morato}
\ead{marcelomnzm@gmail.com}

\author[adUAL]{Igor M. L. Pataro}

\author[adUFBA]{Marcus V. Americano da Costa}

\author[adUFSC]{Julio E. Normey-Rico}

\address[adUFSC]{Renewable Energy Research Group (\emph{GPER}),
  Department of Automation and Systems (\emph{DAS}),\\Federal University
  of Santa Catarina (UFSC), Florian\'opolis, Brazil.}

\address[adUFBA]{Department of Chemical Engineering (\emph{DEQ}),
  Federal University of Bahia (UFBA),\\ 02 Professor Aristides Novis St.,
  Salvador, BA-40210910, Brazil.}

\address[adUAL]{CIESOL, Department of Informatics, University of
  Almer\'ia, Ctra. Sacramento s/n 04120, Almer\'ia, Spain}

\begin{abstract}
The SARS-CoV-2 virus was first registered in Brazil by the end of February, 2020. Since then, the country counts over $70000$ deaths due
to COVID-19, and profound social and economical backlashes are felt all over the country. The current situation is an ongoing health catastrophe, with the majority of hospital beds occupied with COVID-19 patients.  In this paper, we formulate a Nonlinear Model
Predictive Control (NMPC) to plan appropriate social distancing measures (and relaxations) in order to mitigate the pandemic effects, considering the contagion development in Brazil. The NMPC strategy is
designed upon an adapted data-driven Susceptible-Infected-Recovered-Deceased
(SIRD) contagion model, which takes into account the effects of social
distancing. Furthermore, the adapted SIRD model includes time-varying auto-regressive contagion parameters,
which dynamically converge according to the stage of the
pandemic. This new model is identified through a three-layered procedures, with analytical regressions, Least-Squares optimization runs and auto-regressive model fits. The data-driven model is validated and shown to adequately describe the contagion curves over large forecast horizons. In this model, 
control input is defined as finitely parametrized values for social distancing guidelines, which directly affect the transmission and infection rates of the SARS-CoV-2 virus. The NMPC strategy generates piece-wise
constant quarantine guidelines which can be relaxed/strengthen as each week passes. The implementation of the method is pursued through a search mechanism, since the control is finitely parametrized and, thus, there exist a finite number of possible control sequences. Simulation essays are shown to illustrate the results obtained with the proposed closed-loop NMPC strategy, which is able to
mitigate the number of infections and progressively loosen social distancing measures. With respect to an ``open-loop"/no control condition, the number of deaths still could be reduced in up to $30$ \%. The forecast preview an infection peak to September $2$nd, $2020$, which could lead to over $1.5$ million deaths if no coordinate health policy is enacted. The framework serves as guidelines for possible public health policies in Brazil.
\end{abstract}

\begin{keyword}
Nonlinear Model Predictive Control \sep COVID-19 \sep Social isolation
\sep SIRD Model \sep System Identification.
\end{keyword}
\end{frontmatter}


\section{Introduction}
\label{sec1}

 The SARS-CoV-2 virus was first registered in humans in Wuhan, China by December 2019. Since then, it has reached almost all countries around the globe with pandemic proportions and devastating effects. This contagion seems unprecedented and can already be rules as the  definite health crisis of the $21^{\text{rst}}$ century. The SARS-CoV-2 virus causes a severe acute respiratory syndrome, which can become potentially fatal. This virus spreads rapidly and efficiently: by mid-June the virus had already infected over $6$ million people. As of mid-July, over $13$ million COVID-19 cases have been confirmed.
 
 To tackle and extenuate the effects caused by this virus, global scientific efforts have been called out, being urgently necessary \cite{bedford19}. Unfortunately, vaccines take quite some time to be procedure and are, \textit{a priori}, previewed to be ready only by mid-$2021$. To retain the diffuse of this disease, most countries have chosen to adopt social distancing measures (in different levels and with diverse strategies) since March, $2020$ \cite{Adam2020}. This has been explicitly pointed out as the most pertinent control option for the COVID-19 outbreak, including the cases of countries with large social inequalities, as Brazil \cite{baumgartner2020social}. It should be mentioned that the concept underneath social distancing is to impede the saturation of health systems due to large amounts of active COVID-19 infections, which would require treatment at the same time. In this way, when social distancing policies are enacted, the demands for treatment become diluted over time, and the health systems do not have to deal with hospital bed shortages associated with a large peak of active infections.

 Brazil has shown itself as quite a particular case regarding COVID-19 \cite{werneck2020covid}: the country is very large, with $26$ federated states, and each federated state has had autonomy to choose its own health policies, which has lead to different levels of social distancing measures for each state. Furthermore, there has been no coordinated nation-wide public health policy to address the viral spread by the federal government, which is very reluctant to do so, claiming that the negative economic effects are too steep and that social distancing is an erroneous choice \cite{THELANCET20201461,zacchi:hal-02881690}. The country currently ranks as second with respect to numbers of cases and deaths. The expectations disclosed on the recent literature suggest catastrophic scenarios for the next few months \cite{rocha2020expected, morato2020optimal}, which might pursue until mid-$2021$. 

\begin{figure}[htb]
\centering
\begin{subfigure}{.5\textwidth}
  \centering
  \includegraphics[width=1\linewidth]{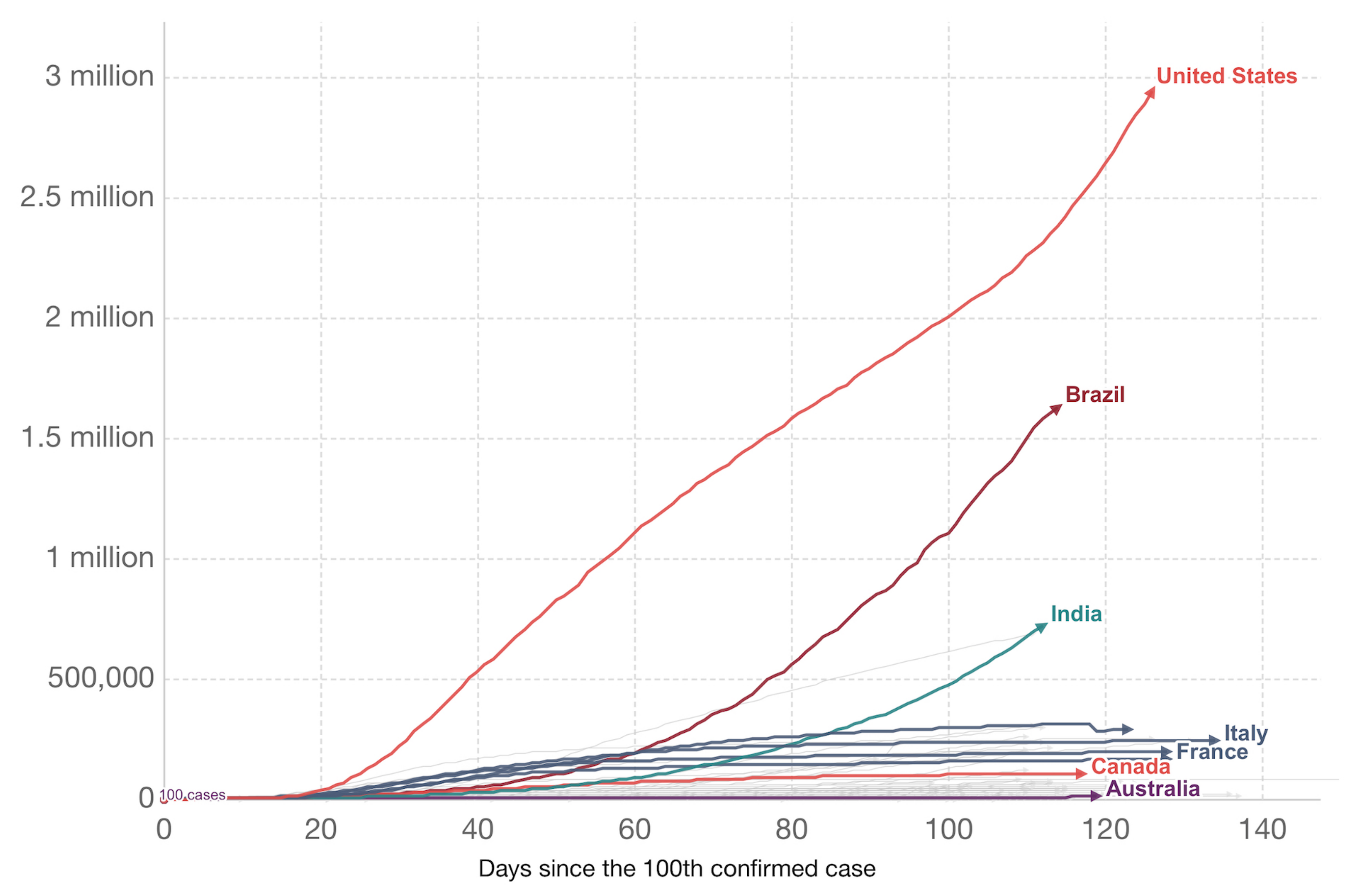}
   \caption{The COVID-19 contagion curve worldwide (in days, since the $100$th confirmed case).}
  \label{Cases_Country}
\end{subfigure}%
\begin{subfigure}{.5\textwidth}
  \centering
  \includegraphics[width=1\linewidth]{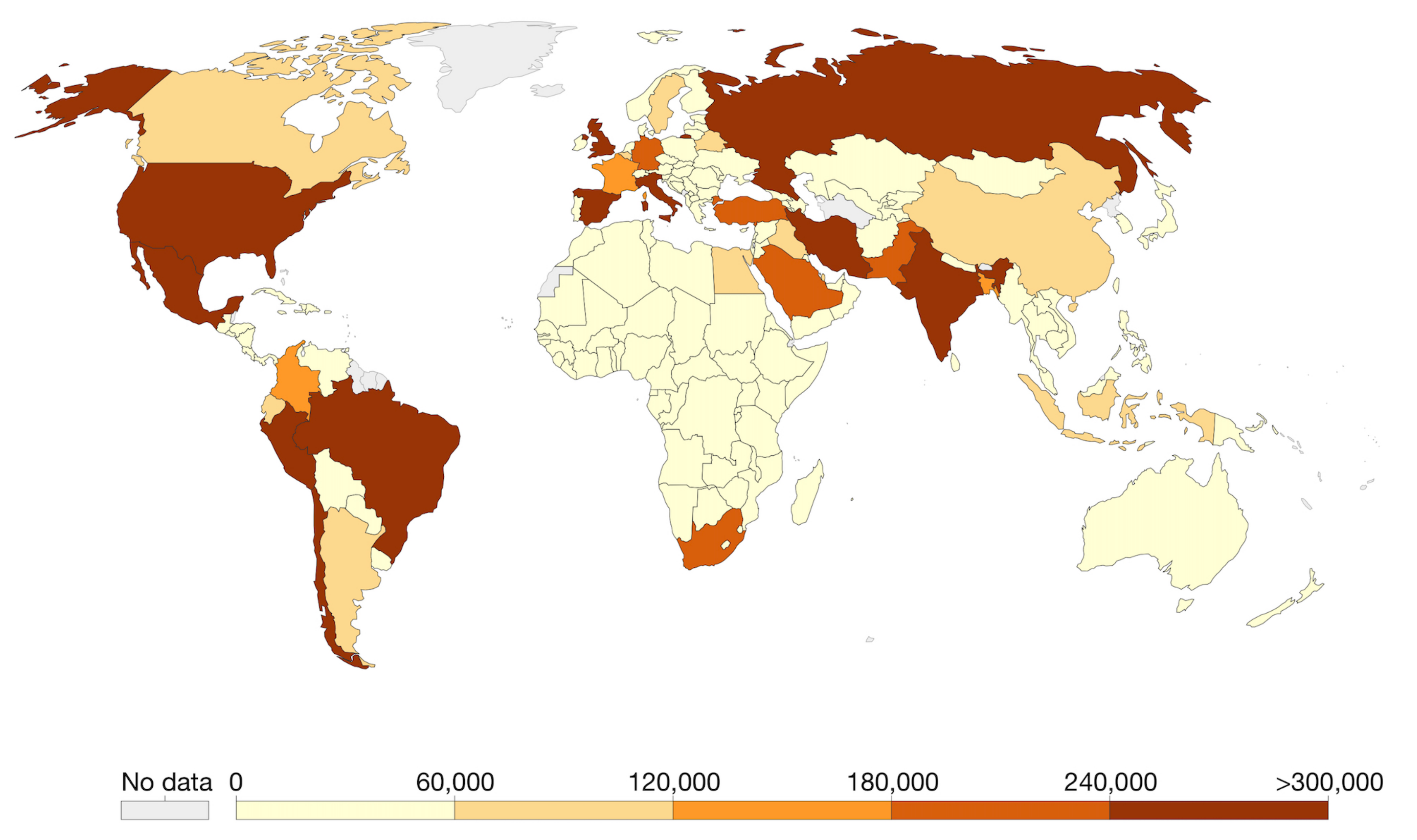}
                \caption{World map COVID-19 case concentration level.}
  \label{Cases_map}
\end{subfigure}
\caption{COVID-19 in the world. Data published by the European Center for Disease Prevention and Control (ECDC),  \cite{owidcoronavirus}.}
\label{DoubleFig}
\end{figure}

\begin{figure}[htb]
	\centering
		\includegraphics[width=0.5\linewidth]{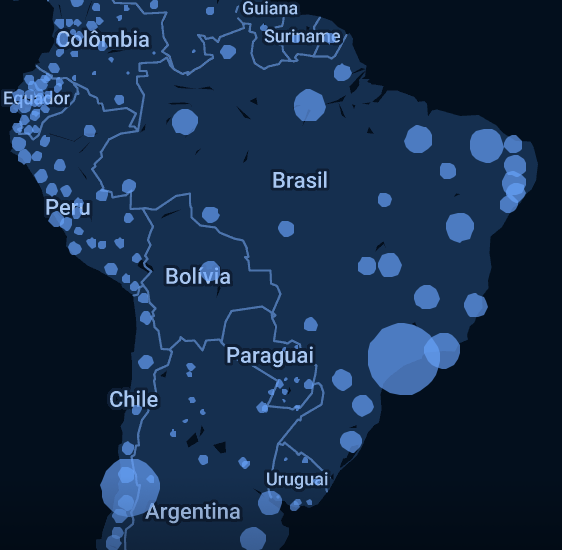}
                \caption{Brazilian map w.r.t. to ``hot-spots" of COVID-19 cases, from Google, dating $30/06/2020$.}
	\label{googlecovidbr}
      \end{figure}

 Regarding this discussion, we consider the COVID-19 contagion data from Brazil in this paper. In order to better illustrate the Brazilian scenario with respect to other countries, Figure \ref{DoubleFig} depicts the evolution of the SARS-CoV-2 contagion curves in different countries of the world, considering the cumulative cases confirmed according to the pandemic period of each country (\ref{Cases_Country}). We note that, as of July $7$th, $2020$, Brazil counts over $1.5$ million confirmed cases of the SARS-CoV-2 virus. The right-side of this Figure (\ref{Cases_map}) maps the concentration level of COVID-19 cases, per country. Complementary, Figure \ref{googlecovidbr} shows the ``hot-spots" of COVID-19 cases in Brazil. The state of S\~ao Paulo concentrates the most number of infections.

      Brazil is currently facing many issues due to the SARS-CoV-2 contagion, despite having a strong universal health system. The current situation is near-collapsing, since the majority of Intense Care Unit (ICU) hospital beds are occupied with COVID-19 patients, all around the country. In addition, the virus is progressing to farthest western cities of the country, away from urban areas, where medical care is somehow less present. It can also be noted that SARS-CoV-2 is currently posing great threat to indigenous communities, such as the Yanomami and Ye'kwana ethnicities\footnote{The Brazilian 
Socioenvironmental Institute (ISA, \textit{Instituto Socioambiental}, see \url{https://www.socioambiental.org/en}) has released a technical note \cite{ferrante2020protect} which warns for the contagion of COVID-19 of up to $40 \, \%$ of Yanomami Indigenous Lands, amid the states of Amazonas and Roraima and a long the border between Brazil and Venezuela, due to the presence of approximately $20000$ illegal mining prospectors. Datasets regarding the COVID-19 spread amid indigenous communities are available in \url{https://covid19.socioambiental.org}.}. Overall, the current situation in the country is very critical.

The first official death due to the SARS-CoV-2 virus in Brazil was registered in March, $17$th $2020$. Anyhow, through inferential statistics, \citet{delatorre2020tracking} acknowledge the fact that community transmission has been ongoing in the state of S\~ao Paulo since the beginning of February (over one month before the first official reports). This points to empirical evidence that the true amount of infected individuals, and possibly registered deaths, are actually very under-reported. Moreover, due to the absence of mass testing, the country is only accounting for COVID-19 patients with moderate to severe symptoms. People with mild or no symptoms are not being accounted for, following guidelines from the federal Ministry of Health. Additionally, we know that the SARS-CoV-2 virus is, for a large number of individuals, an
asymptomatic disease (yet transmissible). For these reasons, the scientific community has been warning for a possibly huge margins of underestimated cases in
Brazil \cite{silva2020bayesian, rocha2020expected, delatorre2020tracking, paixo2020estimation,bastos2020covid19}.  

Regarding this issue, it becomes fundamental to perform coordinated social distancing interventions, as soon as possible. Furthermore, these public health interventions should be put in practice at the correct time and for the correct duration. Well-designed social distance guidelines should help mitigating the contagion and thus avoiding the saturation of the heath systems, while, at the same time, trying to balance social and economic side-effects by releasing/relaxing the quarantine measures as soon as possible. Modeling and identifying the COVID-19 epidemiological process, and designing an optimal controller that addresses this issue are the main focuses of this paper.

Regarding model frameworks for this contagion, recent literature has demonstrated how the SARS-CoV-2 viral contagion dynamics can be very appropriately described by Susceptible-Infected-Recovered-Deceased (SIRD) models \cite{peng2020epidemic,kucharski2020early}. These SIRD models comprise coupled nonlinear differential equations, as originally presented for population dynamics in work of \citet{Kermack1927}. Therefore, the Nonlinear Model Predictive Control (NMPC) \cite{camacho2013model} framework shows itself as a rather convenient approach to guide model-based public health policies, given the fact that it can adequately consider the nonlinear dynamic of the virus spread (through these SIRD equations) together with the effect of lockdown/quarantine measures, introduced as constraints of the optimization problem. Recent literature has indeed shown how SIRD-based NMPCs can be applied to plan COVID-19 social distancing measures in three novel works:
\begin{itemize}
\item \citet{alleman2020covid} consider the application of a NMPC algorithm regarding the data from Belgium. The control input is the actual isolation parameter that is plugged into the SIRD model;
\item \citet{morato2020optimal} consider an optimal On-Off MPC design, which is formulated as mixed-integer problem with dwell-time constraints. Furthermore, the response of the population to social isolation rules with an additional dynamic variable, which is then incorporated to the SIRD dynamics;
  \item \citet{kohler2020robust} applied the technique for the German scenario. The approach considers that the control input is a variable that directly affects the infection and transmission rates (parameters of the SIRD model).
  \end{itemize}

Building upon this previous results, herein we propose an NMPC algorithm formulated on the basis of a SIRD-kind model, which is obtained by the means of an identification algorithm. Following the lines of \citet{morato2020optimal}, we assumed that the control action is a social isolation guideline, which is enacted and passed on to the population. Then, the population responds in some time to these measures, which can be mathematically described as a dynamic social isolation factor. Furthermore, in accordance with the discussion presented by \citet{kohler2020robust}, we also introduce an additional dynamic nonlinear model, which gives the input/output (I/O) relationship between the COVID-19 contagion infection and transmission rates according to the stage of the pandemic, as suggested by \citet{bastos2020covid19}. Complementary, we use auto-regressive equations for the epidemiological parameters of the disease (transmission factor, infection factor, lethality rate) directly related to social distancing as control input. These time-varying regressions are used to provide better long-term forecasts than the regular SIRD equations. Regarding the NMPC framework, the novelty in this paper resides in the formulation of the control input as a finitely parametrized variable, which makes the NMPC implementation persuadable through search mechanisms, which run much faster than the Nonlinear Programming methods from the prior.

Motivated by the previous discussion, the problem of how optimal predictive control can be used to formulate adequate social distancing policies, regarding Brazil, is investigated in this work. The main contribution of our approach comprises the following ingredients:
\begin{itemize}
\item Firstly, an adapted SIRD model is proposed, which incorporates delayed auto-regressive dynamics for the transmission, infection and lethality rates of the SARS-CoV-2 virus, which vary according to the stage of the pandemic. The adapted model also incorporates the dynamic response of the population to a given social distancing guideline, which is the NMPC control input.
\item In order to identify the SIRD model parameter auto-regressive dynamics (for the epidemiological parameters), we use a three-layered identification procedure, which concatenates analytical expansions and Least-Squares optimization procedures. The identified models are validated with regard to the available set of data from Brazil, which includes the social distancing factor observed in the country.
\item The main innovation of the paper is the proposed Parametrized Nonlinear Predictive Control algorithm, which is based on the identified adapted SIRD model with auto-regressive epidemiological parameters. The control input (social distancing guideline) is finitely parametrized over the NMPC prediction horizon, at each sampling instant. Then, an explicit nonlinear programming solver is simulated for all possible input sequences along time. The NMPC solution, then, is simply found through a search mechanism regarding these simulated sequences, which is rather numerically cost-efficient.
\item Finally, we present results considering the application of this NMPC algorithm to the COVID-19 scenario in Brazil. These results are a twofold: a) those that regard the application of the optimal strategy to control the pandemic  \textit{since its beginning}, comparing the simulation results to those seen in practice; and b) applying the control method from $30$th of July onward, aiming to mitigate and revert the current health crisis catastrophe.
\end{itemize}    

This paper is structured as follows. Section \ref{sec2} presents the new SIRD model with auto-regressive epidemiological parameters, the respective identification procedure and validation results. Section \ref{sec3} discusses the proposed NMPC strategy. Section \ref{sec4} depicts the obtained control results, regarding the COVID-19 contagion mitigation for Brazil. General conclusions are drawn in Section \ref{sec5}.
    
\section{Model, Identification and Validation}
\label{sec2}

In this Section, we describe the used modeling framework for the COVID-19 contagion dynamics in Brazil. We build upon the SIRD models for the SARS-CoV-2 virus from works of \citet{peng2020epidemic, kucharski2020early, ndairou2020mathematical}. The model adaptations are done such that the epidemiological parameters are taken as time-varying piece-wise constant functions of the stage of the pandemic. This adaptation is in accordance with recent immunology
results \cite{sun2020understanding, dowd2020demographic, he2020temporal}, which discuss the transmission and reproduction rate of this virus. It is also taken into account two different sampling periods, to include the issue of the incubation period of the virus in human, reportedly, in
average, as $5.1$ days \cite{lauer2020incubation}.

\subsection{SIRD Epidemiological Model}

The SIRD model describes the spread of a given disease with respect to a population split into four non-intersecting classes, which stand for:
\begin{itemize}
    \item The total amount of susceptible individuals, that are prone to contract the disease at a given (discrete) sample of time $k$, denoted through the dynamic variable $S(k)$;
    \item The individuals that are currently infected with the disease (active infections at a given sample of time $k$), denoted through the dynamic variable $I(k)$;
    \item The total amount of recovered individuals, that have already recovered from the disease, from an initial instant $k_0$ until the current sample $k$, denoted through the dynamic variable $R(k)$;
    \item Finally, the total amount of deceased individuals due to a fatal SARS-CoV-2 infection, from an initial instant $k_0$ until the current sample $k$, is denoted $D(k)$.
\end{itemize}   

Due to the evolution of the spread of the disease, the size of each of these classes change over time. Therefore, the total population size \(N\) is the sum of the first three classes as follows:

\begin{eqnarray}
N(k)&=&S(k)+I(k)+R(k)\label{eq:Nconstant} \quad \text{.}
\end{eqnarray}

Since the Brazilian government discloses \textbf{daily samples} of total infections and accumulated deaths, we consider that these discrete-time dynamics samples $k$, given each $T_1 \, = \, 1$ day. Furthermore, to account for the average incubation period of the disease, we consider that the epidemiological parameters vary weekly (each $T_2 \, = \, 7$ days), kept as zero-order-held/piece-wise-constant samples. We will further assess this matter in the sequel.


In the SIRD model, there are three major epidemiological parameters, which express the specific dynamics of the SARS-CoV-2 virus in the population set. The dynamics of these parameters are given in the sparser discrete-time \textbf{weekly samples}, since they change according to the incubation of the virus:
\begin{itemize}
\item The transmission rate parameter \(\beta\) stands for the average number of contacts that are sufficient for transmission of the virus from one individual. According to the detailed classes of individuals, then, it follows that \(T_1\beta(k) I(k)/N(k)\)
  determines the number of contacts that are sufficient for transmission from infected individuals to one susceptible
  individual; and \((T_1\beta (k) I(k)/N(k))S(k)\) determines the number of new cases (per day) due to the amount of \(S(k)\) susceptible individuals (these are the ones ``available for infection''). 
\item The infectiousness Poisson parameter \(\gamma\) denotes the inverse of the period of time a given infected individual  is indeed infectious. Consequently, $\gamma$ affects the rate of recovery (or death) of an infected person. This parameter directly quantifies the amount of individuals that ``leaves'' the infected class, in a given sample.
\item The mortality rate parameter \(\rho\) stands for the
  observed mortality rate of the COVID-19 contagion. We model the amount of deceased individuals due to the SARS-CoV-2 infection following the lines of \citep{keeling2011}: the new number of deaths, at each day, can be accounted for through the following expressions: $\frac{T_1\rho(k)}{1-\rho(k)} \gamma(k) I(k)$
\end{itemize}

Considering these explanations, the ``{\bf SIRD}''
(Susceptible-Infected-Recovered-Dead) model is expressed through the
following nonlinear discrete-time difference equations:
\begin{equation}
    \left\{\begin{array}{rcl}
S(k+1) &=& S(k) - T_1\left(1-\psi(k)\right)\frac{\beta (k) I(k) S(k)}{N(k)} \\[3mm]
      I(k+1) &=& I(k) + T_1\left(1-\psi(k)\right)\frac{\beta (k) I(k)
                 S(k)}{N(k)} - T_1\gamma (k)\frac{I(k)}{1-\rho(k)}
      \\[3mm]
      I_S (k+1) &=& p_{\text{sym}}I(k+1) \\[3mm]
      I_{c} (k+1) &=& \left(I(k+1) + R(k+1) + D(k+1) \right) \\[3mm]
      R(k+1) &=& R(k) + T_1\gamma (k)I(k) \\[3mm]
      D(k+1) &=& D(k) + T_1\frac{\rho(k)}{1-\rho(k)}\gamma (k) I_S (k)
             \\[3mm]\end{array}\right.\;\;\;\textrm{\bf [SIRD]} \quad \text{,}
\label{eqSIRDmodel}
\end{equation}
where $I_S$ denotes the portion of the infected individuals which in fact display symptoms. This ``symptomatic" class has been accounted for in previous papers \cite{bastos2020modeling,morato2020optimal}. We note that only these symptomatic individuals will require possible hospitalization and may die. The remainder $(1-p_{\text{sym}})I$ are asymptomatic or lightly-symptomatic, which do transmit the virus but do not die or require hospitalization. The class of recovered individuals considers immunized people, encompassing those that displayed symptoms and those that did not. The symptomatic ration parameter $p_{\text{sym}}$ is constant and borrowed from previous papers.

The cumulative number of cases $I_c$ denotes the total number of people that have been infected by the SARS-CoV-2 virus until a given day. In fact, this cumulative variable is equivalent to those that are currently infected $I$, summed with those that have recovered $R$ and those that have died $D$.

We must also stress that, in the SIRD model equations given above, parameter $\psi$ represents a transmission rate mitigation factor. This factor accounts for the observed social isolation factor with the population set $N$. It follows that $\psi \, = \, 0$ denotes the situation where the whole population set has sustained social interactions. As discussed in previous papers \citep{keeling2011,bastos2020covid19}, there exists some "natural" $\psi \, = \, \underline{\psi}$ factor, which stands for a normal/nominal conditions, still with ``no control'' of the viral spread (with no social isolation guidelines, this kind of situation has been observed in Brazil in the first weeks of the contagion, end of February, beginning of March). In contrast, $\psi \, = \, 1$ represents a complete lockdown scenario, where there are no more social interactions (this conditions has not been seen and is, in practice, unattainable). In practice, there is some maximal social isolation factor $\psi \, = \, \overline{\psi}$ that can be put in practice. In this paper, we consider the values for social isolation in Brazil from the work of \citet{bastos2020covid19}, which discloses weekly estimates for $\psi$.

Another essential information in epidemiology theory is the basic reproduction number, usually denoted by $R_0$. This factor is able to measure the average potential transmissibility of the disease. In practice, this value is fixed/constant and inherent to a given disease. Through the sequel, a time-varying version of this basic reproduction number is considered, denoted $R_t$, which represents how many COVID-19 cases could be expected to be generated due to a single primary case, in a population for which all individuals are susceptible. From a control viewpoint, $R_t$ represents the epidemic spread velocity: if $R_t> 1$, the infection is spreading and the number of infected people increases along time (this typically happens at the beginning of the epidemic), otherwise, if $R_t<1$, it means that more individuals ``leave'' from the infected class, either recovering or dying, and, thereby, the epidemic ceases. The reproduction number $R_0$ is affected by different factors, including immunology of the virus, biological characteristics, but also governments policies to control the number of susceptible population. Since the epidemic parameters change along the time, we process by considering the effective reproduction number $R_0$ also as a dynamic variable.

The underlying assumption used to calculate $R_t$, is that, at the beginning of the pandemic, $S\approx N$. Considering parameters $\beta$, $\gamma$, $\rho$ and $\psi$ from the SIRD model Eq. \eqref{eqSIRDmodel}, $R_t$ can be roughly given by:
\begin{eqnarray} 
\label{R0eq} R_t(k) \approx \frac{(1-\psi (k))\beta (k)(1-\rho (k))}{\gamma
  (k)} \quad \text{.}
\end{eqnarray}

\begin{remark}
  Regarding the SIRD model given in Eq. \eqref{eqSIRDmodel}, it
  follows that $N(0) = N_0$ is the initial population
 size (prior to the viral infection). Furthermore, we stress that, in SIRD-kind models, $I(k)$
 represents the \textbf{active} infections at a given
 moment, while $D(k)$ represents the \textbf{total} amount of deaths until this
 given moment; for this reason, it follows that $D(k+1) - D(k)$ is
 proportionally dependent to $I(k)$.
\end{remark}

 \begin{remark}
   We note that we are not able to use more ``complex'' descriptions
   of the COVID-19 contagion in Brazil, such
   as the ``SIDARTHE'' model used by \citet{kohler2020robust} (that
   splits the infections into (symptomatic, asymptomatic) detected,
   undetected, recovered, threatened and extinct) because we have
   insufficient amount of data. The Ministry of Health only disclosed
   the total amount of infections ($I(k) + D(k) + R(k)$) and the total
   amount of deaths ($D(k)$) at each day. Due to the
   absence of mass (sampled) testing, there is no data regarding
   detected asymptomatic individuals, for instance, as it is available
   in Germany (where \citep{kohler2020robust} originate). To choose a
   more complex model may only decrease the truthfulness/validity of
   the identification results, since the parameters can only represent
   a singular combination that matches the identification datasets and
   cannot be used for forecasting/prediction purposes.
 \end{remark}

\begin{figure}[htb]
	\centering
		\includegraphics[width=\linewidth]{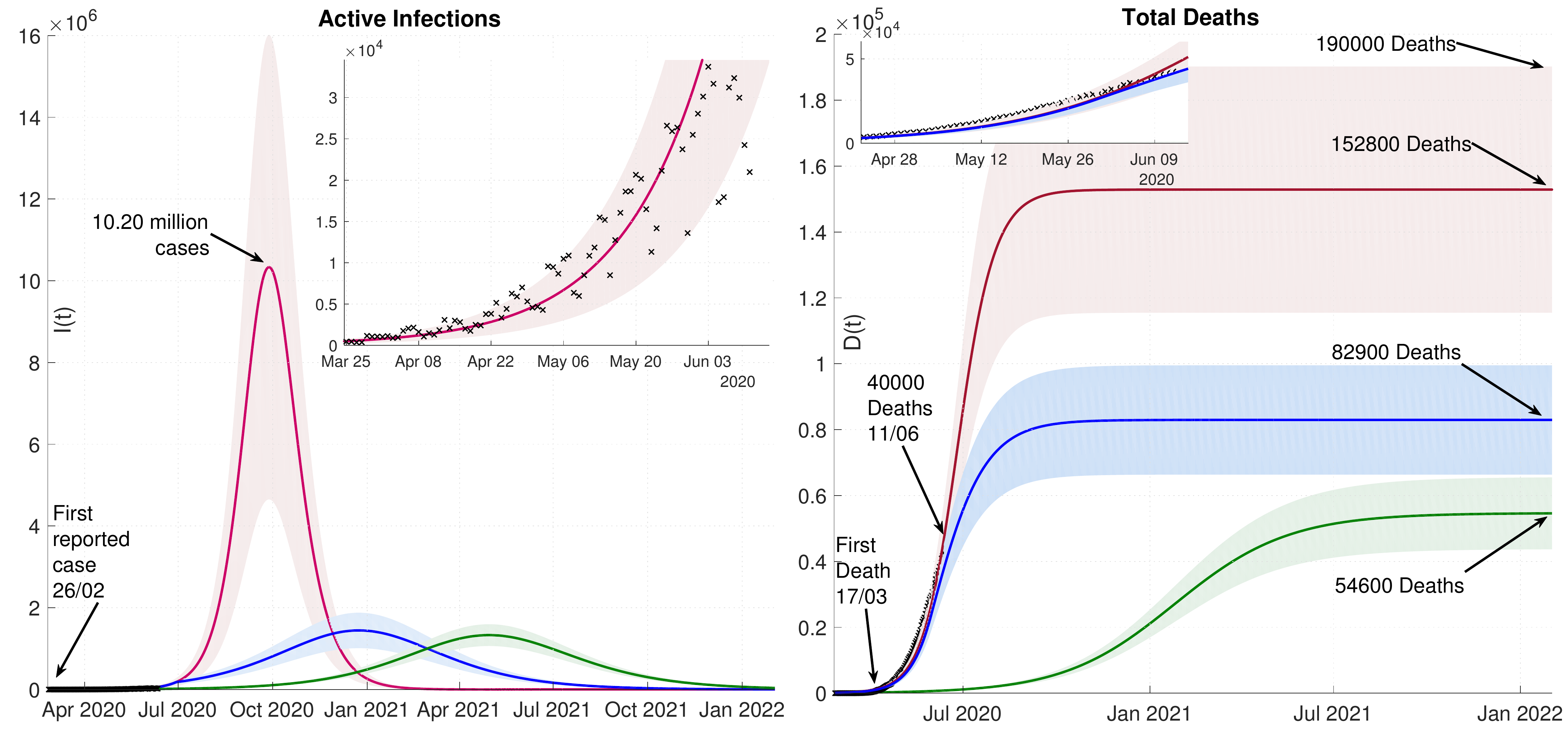}
                \caption{Long-term SIRD Forecasts for Brazil (from $11/06/2020$). Consolidated datasets ($I(t)$ and $D(t)$): a)
                  No-Control Situation (\textcolor{red}{$\mathbf{-}$}), b)
                  Hard social distancing ($\psi \, = \, 0.6$) from $11/06/2020$ onwards
                  (\textcolor{blue}{$\mathbf{-}$}), c) $\psi = 0.6$ Hard social distancing applied from March
                  (\textcolor{green}{$\mathbf{-}$}) and d) Real
                  data ($\times$). Shades represent total
                  variation over a $95\%$ confidence interval, solid lines represent mean values.}
                \label{forecasts}
\end{figure}

 To illustrate the dynamics of a contagion like the pandemic outbreak
 of COVID-19, in Figure \ref{forecasts} we present long-term forecast using a regular SIRD model
 with constant epidemiological parameters, following the methodology
 presented by \citet{bastos2020covid19}. These predictions were
 computed on $11/06/2020$ and are here shown to demonstrate the
 qualitative evolution of the $I$ and $D$ curves. The active
 infections curve $I$ shows a increase-peak-decrease characteristic,
 while the total number of deaths $D$ shows an asymptotic behavior to some steady-state condition.  We note that, as of $30/06$, the number of infections had already significantly increased, which means that the most recent forecasts preview an even worse catastrophe. 

 \subsection{Model Extensions}
 \label{modelextensionsec}

 In this paper, the SIRD dynamics from Eq. \eqref{eqSIRDmodel} are adapted. These adaptations are included for three main reasons:
 \begin{enumerate}
 \item In accordance with the discussions presented by \citet{morato2020optimal},
   we understand that it is imperative to embed to the SIRD model how
   the population responds to public health policies, that is, to
   incorporate the dynamics and effects of social distancing. When a
   government enacts some social distancing policy, the population
   takes some time to adapt to it and to, in fact, exhibit the
   expected social distancing factor $\psi$. This is, to take $\psi$ as a time-varying dynamic map of the enacted social distancing guidelines (which will, later on, be defined by an optimal controller).
 \item Many results seen in the literature \cite{bastos2020modeling,
     ImperialCollegeNew, ndairou2020mathematical} consider constant
   parameter values for $\gamma$, $\beta$ and $\rho$, understanding
   that these factor are inherent to the disease. Anyhow, recent
   results \cite{bastos2020covid19, kucharski2020early} indicate that these parameters are indeed time-varying, achieve steady-state values
   at ending stages of the pandemic. It is reasonable that when more
   active infections occur, the virus tends to spread more
   efficiently, for instance. Moreover, the mortality rate should stabilize after
   the infection curve has decreased. To incorporate this issue in the
   model, auto-regressive moving-average time-varying dynamics for the viral
   parameters are proposed, which stabilize according to stage of the pandemic. Furthermore, by doing so, the
   inherent incubation characteristic of the virus is also taken into
   account, since the viral transmission, lethality and infection
   rates should vary with dynamics coherent with the average
   incubation period.
 \item Recent papers provide forecasts
   \cite{bastos2020modeling, ImperialCollegeNew} for the evolution of
   the virus assuming that the epidemiological parameters remain
   constant along the prediction horizon. This is not reasonable, due
   to the fact that such parameters are inherently time-varying, given
   according to the exhibited pandemic level. Thus, if parameters are
   held constant for forecasts, these forecasts are only qualitative
   and allow short-term conclusions. Since dynamic models
   for the epidemiological parameters are proposed, make more
   coherent long-term forecasts can be derived. Of course, it must be stressed that these
   forecasts will still be qualitative, since one cannot account for
   perfect accurateness regarding the number of infection and deaths
   due to the absence of mass testing in Brazil. Furthermore, the effect of future unpredicted phenomena cannot be accounted for, such as the
   early development of an effective vaccine, which would certainly
   make the infections drop largely.
 \end{enumerate}

These model adaptations are discussed individually in the sequel.
 
\subsubsection{Social Distancing Guidelines / The Control Input}

 In order to design and synthesize effective control strategies for
 social distancing (public) policies, to be oriented to the population
 by local governments, the social distancing factor $\psi$ is further
 exploited.
 
 In what concerns the available data from Brazil (that is used for identification of the model parameters), the social distancing factor $\psi$ is a known variable, given in weekly piece-wise constant samples. Later on, regarding the control procedure (Section \ref{sec4}), this isolation factor will
 vary according to the enacted social distancing policy $u$, as defined by
 a nonlinear optimal predictive control algorithm.

The differential equation that models the response of the susceptible population to quarantine guidelines is taken as suggests \cite{morato2020optimal}, this is:
\begin{equation}
   \left\{ \begin{array}{rcl}
             \psi (k+1) & = & \psi (k) +
                                T_2\varrho_{\psi}\left(K_\psi(k)u(k)
                                - \psi(k)\right) \end{array}\right. \quad \text{,}
  \label{TheControlPsi}
\end{equation}
where $u(k)$ is the actual control input, the guideline that defines the social isolation factor goal, as regulated by the government (this signal will be later on determined by the proposed optimal controller), $\varrho_{\psi} \, \, = \,\, 0.4317 \,\, \rm{day^{-1}}$ is a settling-time parameter, which is related to the average time the population takes to respond to the enacted social isolation measures, and
$K_\psi$ is a time-varying stating gain relationship between $\psi$ and $u$.

As recommend \citet{morato2020optimal}, we assume that when more infected cases have been reported, and when the hospital bed occupation surpasses $70 \%$, the population will be more prone to follow the social distancing guidelines, with larger values for the gain relationship $K_\psi$:
\begin{eqnarray}
  \label{TheKPsiEq}
  K_\psi(k) &=& \max\left\{1 \,\text{,} \,\frac{p_{\text{sym}}I(k)}{0.7n_{ICU}} \right\} \quad \text{,}
\end{eqnarray}
being $n_{ICU}$ the total number of ICU beds in the country. We recall that $p_{\text{sym}}$ is a parameter which gives the amount of infected individuals that in fact display symptoms and may need to be hospitalized ($I_s \, = \, p_{\text{sym}}I$). 

\begin{remark}
  We note that Brazil has, as of February, $45,848$ ICU beds. This number is estimated to increase in up to $80 \, \rm{\%}$ with field hospitals that were built specifically for the COVID-19 contagion (i.e. $n_{ICU} \, = \, 82,526$). The percentage of symptomatic individuals is taken as $16 \, \rm{\%}$, according to the suggestions of \citet{bastos2020modeling}. This value is coherent with the available information regarding this virus, concerning multiple countries \cite{lima2020information}. The percentage of symptomatic considered groups the infections with severe/acute symptoms, which will, indeed, most possibly require treatment.
\end{remark}

\begin{remark}
In practice, Eq. \eqref{TheControlPsi} is bounded to the minimal and maximal values for the social distancing factor $\underline{\psi}$ and $\overline{\psi}$, respectively. These values are the same are that used as saturation constraints on $u$, as discussed in the sequel.
\end{remark}

\begin{figure}[htb]
	\centering
		\includegraphics[width=\linewidth]{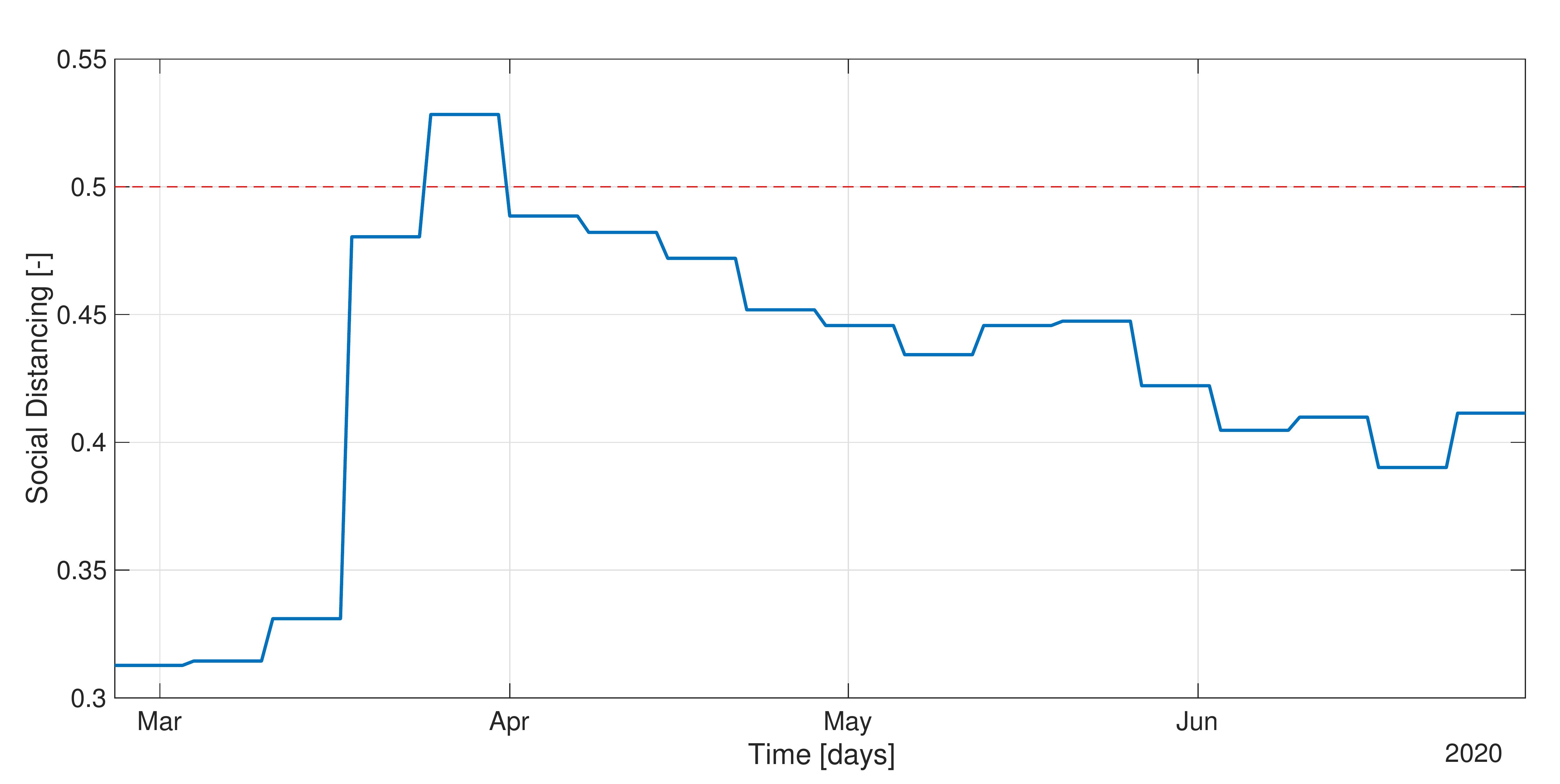}
                \caption{Social distancing factor in Brazil from February $26$th, $2020$ to June, $30$th $2020$ \cite{InLoco}.}
	\label{psi_br}
\end{figure}

We proceed by considering that $u$ is a finitely parametrized control input. This is: the enacted social distancing guidelines can only be given within a set of pre-defined values. This approach is coherent with possible ways to enforce and put in practice social distancing measures. Pursuing this matter, the actual control signal, that is to be defined by the proposed optimal automatic controller,
must abide to a piece-wise constant characteristic with a possible increase/decrease of $5\%$ of isolation per week. This percentage is the average increase of social isolation, in the beginning of the pandemic in Brazil, when social distancing measures were strengthened significantly over a small time period. The values for social isolation $\psi(k)$ in Brazil have been estimated by \citet{InLoco} and are presented in Figure \ref{psi_br}.

Moreover, $u$ is considered to be defined within the admissibility set $[\underline{\psi} \, , \, \overline{\psi}]$. As gives Figure \ref{psi_br}, which shows the observed social isolation factor $\psi$ in Brazil, the ``natural" isolation factor is of $\underline{\psi} \, = \, 0.3$ (which stands for a situation of no-isolation guidelines). Furthermore,  $\overline{\psi}$ is the maximal attainable
isolation factor for the country; the maximal observed value in Brazil is of roughly $53 \, \%$ (see Figure \ref{psi_br}). Anyhow, coordinated ``lockdown" measures were not forcefully enacted. For this matter, we consider $\overline{\psi} \,=\,
0.7$, which would represent that the population can be restricted to, at most, a $70\%$ reduction on the level of social interactions. As reported by \citet{bastos2020covid19}, it seems unreasonable to consider larger values for social isolation in the country, due to multiple reasons (hunger, social inequalities, laboring
necessities, lack of financial aid from the government for people to
stay home, \textit{etc.}).

\begin{remark}
  It must noted that the SIRD model considers some ``natural" isolation factor (which is the lower bound on $u$). As seen in Figure \ref{psi_br}, the social isolation factor of $30\,\rm{\%}$ is the observed lower bound for $\psi$ seen in Brazil.
\end{remark}

Mathematically, these constraints are expressed through the following Equation set:
\begin{eqnarray}
  \label{UConstraints} \left\{\begin{array}{rcl}
  u(k) &=& u(k-1) + \delta u(k) \\
  \delta u (k)  &=& \{-0.05 \text{ or } 0 \text{ or } 0.05\}\\
  u(k) &\in& \left[\begin{array}{ccccccc}0.30 & 0.35& 0.40& \dots &
                                                                 0.60
                       &0.65&
                              0.70  \end{array}\right] \end{array}\right. \quad \text{.}  
\end{eqnarray}

This constraints given in Eq. \eqref{UConstraints} are, in fact, very
interesting from an implementation viewpoint, as the government could
convert the finitely parametrized values for $u$ into actual
practicable measures, as illustrates Table
\ref{TheTableUParametre}. We remark that this Table is only
illustrative. Epidemiologists and viral specialists should be the ones
to formally discuss the actual implemented measures that ensure the
social distancing factor guideline given by $u$.

\begin{table}[htbp]
    \caption{\label{TheTableUParametre} Illustrative Example of
      Finitely Parametrized Social Distancing Measures.}
    \centering
	\begin{tabularx}{\textwidth}{|c | X | c |}
		\hline \hline
        Control Signal / Social Distancing Guideline ($u$) & Implemented
                                                       measures &
                                                             Infection
                                                                  Risk \\\hline
          $u \, = \, 0.3$ & No public health emergency. All economy
                          sectors can return to their normal activities. & Controlled Contagion \\ \hline
          $\vdots$ & $\vdots$ & \\\hline
          $u \, = \, 0.35$ & Low restriction levels. Use of masks to go
          outside. Public transport functioning. Limited opening of
          shops and small public spaces. & Low Risk \\ \hline
            $\vdots$         & $\vdots$ & \\\hline
           $u \, = \, 0.4$ & Moderate restrictions. Use of masks to go
          outside. Closed public spaces. Restricted openings only.  & Moderate Risk \\ \hline
            $\vdots$         & $\vdots$ & \\\hline
            $u \, = \, 0.45$ & Very restrictive policies. Reduced public
            transport. Urge for people to stay home at all times. Very
            restrictive openings only. & High Risk \\ \hline
              $\vdots$       & $\vdots$ & \\\hline
          $u \, = \, 0.7$ & Severe restrictive policies. No public
            transport. Urge for people to stay home at all times. Only
            basic services may open, with reduced capacities.  & Very High Risk \\ \hline
    \end{tabularx}
\end{table}

\subsubsection{Dynamic Epidemiological Parameter Models}

The second main modification of the original SIRD model is to consider
dynamic models for the epidemiological parameters of the SARS-CoV-2
virus, $\gamma$, $\beta$ and $\rho$. These models are taken as auto-regressive moving average functions, which converge as the pandemic progresses. Therefore, the following dynamics are considered:
\begin{eqnarray}
  \label{TheDynamicsEpModel}
  \left\{\begin{array}{rcl}
\beta (k) &=& f_\beta \left(\beta (k-1)\, \dots, \beta (k-n_\beta)\right) \\
           \gamma (k) &=& f_\gamma \left(\gamma (k-1)\, \dots, \gamma (k-n_\gamma)\right) \\
           \rho (k) &=& f_\rho \left(\rho (k-1)\, \dots, \rho (k-n_\rho)\right) \\ \end{array}\right. \quad \text{.}
  \end{eqnarray}

  The models given in Eq. \eqref{TheDynamicsEpModel} are possibly delayed and auto-regressive. Anyhow, despite the
  parameters $\beta$, $\gamma$ and $\rho$ being time-varying, the
  model functions $f_\beta$, $f_\gamma$ and $f_\rho$ are constant. The order and number of regressions are found through an optimization procedure, which is further detailed in Section \ref{Identifsec}.

  \subsubsection{The Complete COVID-19 Model}
  
\begin{figure}[htb]
	\centering
		\includegraphics[width=\linewidth]{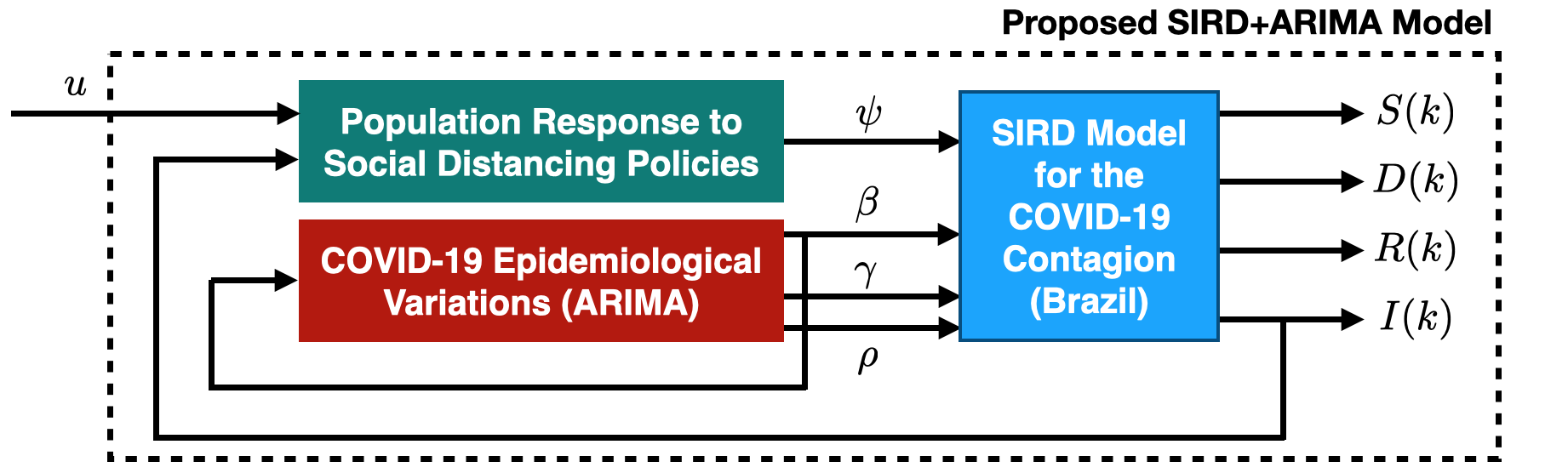}
                \caption{The ``SIRD$+$ARIMA" model for COVID-19 in Brazil.}
	\label{modeltotal}
\end{figure}

  The complete model used in this work to describe the COVID-19
  contagion outbreak in Brazil is illustrated through the
  block-diagram of Figure \ref{modeltotal}. We note that the
  ``COVID-19 Epidemiological Parameter Models'', represented through
  Eq. \eqref{TheDynamicsEpModel}, enable the complete model to offer
  long-term predictions with more accurateness, as discussed in the
  beginning of this Section.  We denote as the ``SIRD$+$ARIMA" model, henceforth, the cascade of the population response from Eq. \eqref{TheControlPsi} to the SIRD model, with auto-regressive time-varying epidemiological parameters as give Eq. \eqref{TheDynamicsEpModel}.

\subsection{Identification Procedure}
\label{Identifsec}

Recent numerical algorithms have been applied to estimate the model parameters of the COVID-19 pandemic \cite{bastos2020covid19,SARKAR,SUN2020,KENNEDY20,Caccavo2020,Oliveira2020,Villaverde2020estimating}. The majority of these papers indicates that SIRD models provide the best model-data fits.

We note, anyhow, that the SIRD model offers three degrees-of-freedom at each instant $k$ (i.e. $\beta(k)$, $\gamma (k)$ and $\rho(k)$), as gives Equation \ref{eqSIRDmodel}. This means that different instantaneous combinations of these parameters can yield the same numerical values for $S(k)$, $I(k)$, $R(k)$ and $D(k)$. Therefore, although mathematical and graphical criteria have been used to validate these dynamic models when compared to real data, the estimated values for these parameters should be coherent with biological characteristics of this viral pandemic. An indicative of badly adjusted SIRD model parameters is the effective reproduction number of the disease $R_0$, which should naturally decrease to a threshold smaller than one as the pandemic ceases.

Thus, in order to identify a SIRD model that is indeed able to describe the pandemic contagion accurately (which is especially important for forecast and control purposes), we proposed a two step identification method to determine the time-varying dynamics for $\beta(k)$, $\gamma(k)$ and $\rho(k)$, with respect to the dynamic parameter models in Eq. \eqref{TheDynamicsEpModel}. 

Regarding datasets, we must comment that the Brazilian Ministry of Health discloses, daily, the values for the cumulative number of confirmed cases $I_{c}(k)$ and the cumulative number of deaths $D(k)$. Furthermore, we recall that, through the sequel, the values of the observed average social isolation in the country, $\psi(k)$, are assumed to be known (as given in Figure \ref{psi_br}). The symptomatic percentage is assumed as constant along the prediction horizons.

Then, we proceed by depicting the used identification procedure, which is performed in three consecutive steps/layers, as follows:
\begin{itemize}
\item The first step resides in analytically solving the SIRD regressions from Equation \ref{eqSIRDmodel} for a fixed interval of points (days).
\item Then, the derived values from the analytical solutions are passed as initial conditions to an optimization layer, which solves a constrained Ordinary Least Square minimization problem, aiming to adjust the parameter values to pre-defined (biologically coherent) sets, so that the identified model yields small error with adequate parameter choices. The output of this second step stands for the time-series vectors regarding the SIRD parameters.
\item Finally, these time-series are used to fit auto-regressive models, via Least Squares, as gives Eq. \eqref{TheDynamicsEpModel}.
\end{itemize}

By following this three-layered procedure, the estimated parameter equations (Eq. \eqref{TheDynamicsEpModel}) are found in accordance with biological conditions, which are described as the pre-defined optimization sets. We note that we embed to the optimization layer sets that are also in accordance with previous results for SIRD model estimations for Brazil \cite{bastos2020modeling,morato2020optimal}.

\begin{figure}[htb]
	\centering
		\includegraphics[width=1\linewidth]{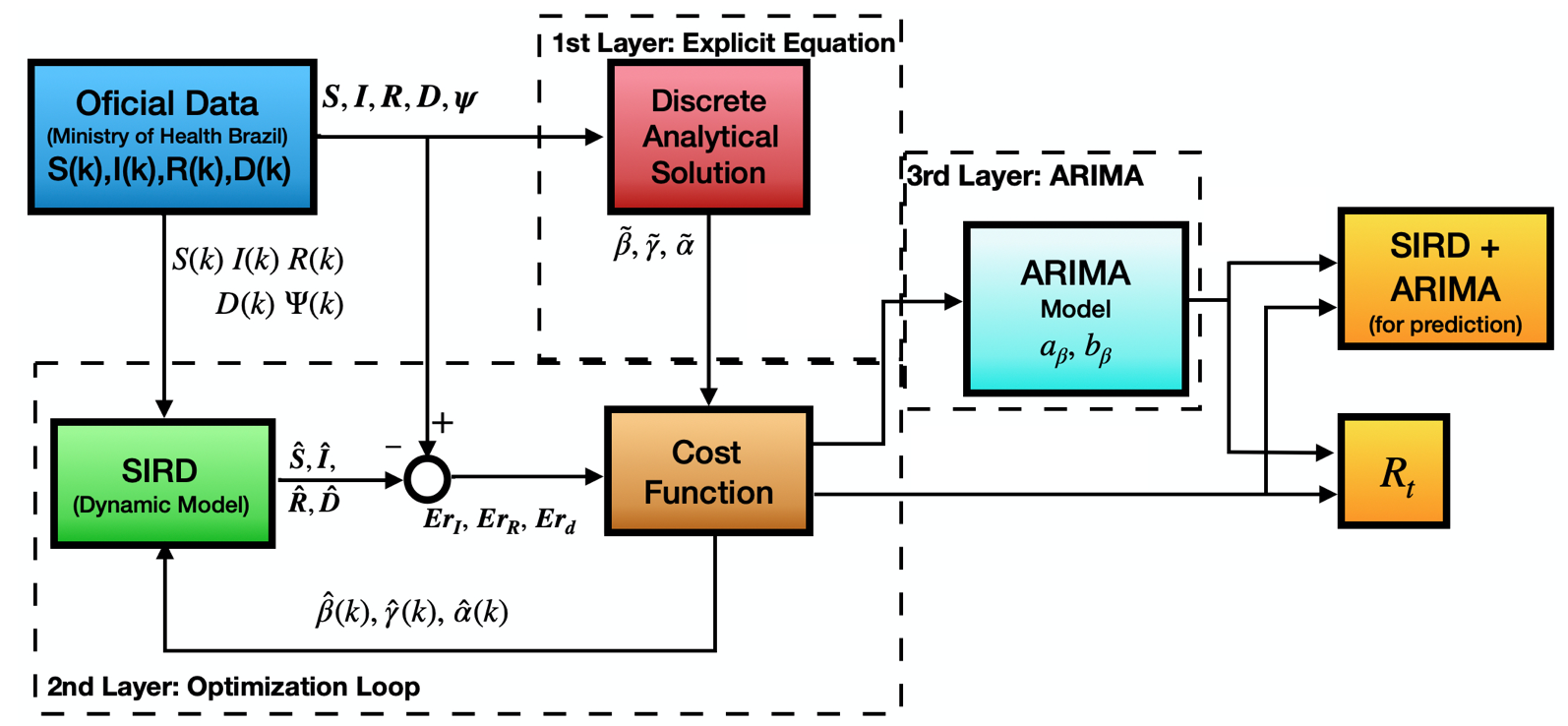}
                \caption{Parameter Identification Procedure.}
	\label{ident_algorithm}
\end{figure}

The complete identification procedure used to estimate the SIRD model dynamics follows the lines illustrated in Figure \ref{ident_algorithm}.

\subsection{First Layer: Analytical Solutions}

We note that, in order to simplifying the formulation of the optimization algorithm, the difference Equations for $I(k+1)$ and $D(k+1)$ are modified in order to decouple parameters related to the number of fatal and related to the number of recovered individuals, with regard to $I(k)$. This is:
\begin{eqnarray} 
        I(k+1) &=& I(k) + T_1\left(1-\psi(k)\right)\frac{\beta (k) I(k)S(k)}{N(k)} - T_1\gamma (k)I(k) - T_1\alpha (k)I(k) \\[3mm]
        D(k+1) &=& D(k) + T_1\alpha(k) I (k)
\end{eqnarray}
\noindent where $\alpha(k) = \rho(k) \frac{\gamma(k)}{(1-\rho(k))}$. Therefore, instead of identifying $\beta$, $\gamma$ and $\rho$, we pursue the identification of $\beta$, $\gamma$ and $\alpha$, and, then, $\rho$ is computed as follows:
\begin{eqnarray} 
\label{rhoalfarelationship}
        \rho (k) &=& \frac{\frac{\alpha(k)}{\gamma (k)}}{1+\frac{\alpha(k)}{\gamma (k)}}\quad \text{.}
\end{eqnarray}

The identification procedure starts by collecting the available datasets (regarding the cumulative number of COVID-19 cases $I_c$ and deaths $D$) inside a fixed interval $k \, \in \,[k_{i}, k_{f}]$. Then, the first layer computes ``exact", analytical values for the epidemiological parameters, denoted $\tilde{\beta}$, $\tilde{\gamma}$ and $\tilde{\alpha}$ according to the following discrete analytical expansions:
\begin{eqnarray}
\tilde{\gamma} &=& \frac{R(k_f)-R(k_i)}{\sum_{i=k_i}^{i=k_f}I(i)} \, \text{,}\\[3mm]
\tilde{\alpha} &=& \frac{D(k_f)-D(k_i)}{\sum_{i=k_i}^{i=k_f}I(i)} \, \text{,}\\[3mm]
\tilde{\beta}&=&\frac{1}{(1-\psi(k_f))} \frac{I(k_{f})-I(k_{i})+(\tilde{\alpha}+\tilde{\gamma})\sum_{i=k_{i}}^{i=k_{f}} I(i)}{\sum_{i=k_{i}}^{i=k_{f}}S(i)I(i)/N(i)} \, \text{.}
\end{eqnarray}

\subsection{Second Layer: Ordinary Least Squares Optimization}

It is assumed that the used datasets might be corrupted by series of issues, such as cases that are not reported on given day $k$ and simply accounted for on the following days, or sub-reported cases, as discussed by \citet{bastos2020covid19}. Therefore, we adjust these parameters through the optimization layer, in order to improve the reliability of the identification.

Once again, the available data from the same interval discrete $[k_{i}, k_{f}]$ is used. The optimization procedure minimizes a constrained Ordinary Least Squares problem, whose solution comprises the following vectors: $\boldsymbol{S}$, $\boldsymbol{I}$, $\boldsymbol{R}$ and $\boldsymbol{D}$. These vectors comprise the model-based outputs within the considered discrete interval. The decision variables for the optimization procedure are the degrees-of-freedom of the SIRD model, in this layer denoted $\hat{\beta}$, $\hat{\gamma}$, $\hat{\alpha}$.  The Ordinary Least Square criterion ensures that the optimization minimizes the error between the real data and the estimated values from the SIRD model, by choosing coherent values for the decision variables.

The quadratic model-data error variables terms used in the optimization layer are the following:
\begin{eqnarray} \label{eq:err}
Er_{I}(i) &=& (I(i) - \hat{I}(i,\hat{\beta} (1-\psi),\hat{\gamma},\hat{\alpha}))^2\text{,} \\ \label{eq:err2}
Er_{R}(i) &=& (R(i) - \hat{R}(i,\hat{\beta}(1-\psi),\hat{\gamma},\hat{\alpha}))^2 \text{,}\\ \label{eq:err3}
Er_{D}(i) &=& (D(i) - \hat{D}(i,\hat{\beta}(1-\psi),\hat{\gamma},\hat{\alpha}))^2 \text{,}
\end{eqnarray}
\noindent for which the variables $\hat{I}, \hat{R}, \hat{D}$ are estimated according to the SIRD model equations. The complete optimization problem is formulated as follows:

\begin{eqnarray}
\nonumber
    &&\underset{\beta,\gamma,\alpha}{\min}  J =  \sum_{i=k_i}^{i=k_f} \left(w_1Er_I(i) + w_2 Er_R(i) + w_3 Er_D(i)\right) \text{,} \\ \label{eq:opt}
    \text{s.t.:} && \underline{\delta} \tilde{\beta}(1-\psi(i)) \leq \beta (1-\psi(i)) \leq \overline{\delta} \tilde{\beta}(1-\psi(i)) \text{,}\\ \nonumber
    && \underline{\delta}\tilde{\gamma} \leq \gamma \leq \overline{\delta}\tilde{\gamma} \text{,}\\ \nonumber
&& \underline{\delta}\tilde{\alpha} \leq \alpha \leq \overline{\delta} \tilde{\alpha}\text{.} \\ \nonumber 
&& \begin{bmatrix}
\beta_0\\ 
\gamma_0\\ 
\alpha_0 \end{bmatrix} = \begin{bmatrix}
\tilde{\beta} \\ 
 \tilde{\gamma} \\ 
 \tilde{\alpha}
\end{bmatrix} \text{.} 
\end{eqnarray}
\noindent wherein $\underline{\delta}$ and $\overline{\delta}$ are uncertainty interval margins used to define the lower and upper bound of each decision variable on the optimization problem, $\beta_0$, $\gamma_0$ and $\alpha_0$ are the initial conditions of the optimization problem and $w_1$, $w_2$ and $w_3$ are taken as positive weighting values (tuning parameters), used to normalize the magnitude order of the total cost and adjust the model fit with respect to variables $Er_I$, $Er_R$ and $Er_D$. 

With respect to the discussion regarding the average incubation period of the SARS-CoV-2 virus, the optimization procedure compute piece-wise constant values for $\beta(k)$, $\gamma(k)$ and $\alpha(k)$ each $T_2 \, = \, 7$ days. This is also done to improve the model-data fitting results, as suggest \citet{morato2020optimal}. 


Considering the application of this second layer to a complete data-set, with new values for the epidemiological parameter each $T_2$ days, the output of this layer stands for the epidemiological time-series vectors denoted $\beta_{opt}$, $\gamma_{opt}$ and $\alpha_{opt}$. These time-series are, then, used to fit the auto-regressive models in Eq. \eqref{TheDynamicsEpModel}, in the third layer. We note that these SIRD parameter dynamics are the ones that can be can be used for forecasting and control purposes (and also to calculate the effective reproduction number $R_t$). 


\subsection{Third Layer: ARIMA fits}

The third layer of the identification procedure resides in fitting an auto-regressive model to the time-series derived from the optimization $\beta_{opt}$, $\gamma_{opt}$ and $\rho_{opt}$ (note that $\rho_{opt}$ is given as a function of $\beta_{opt}$, $\gamma_{opt}$ and $\alpha_{opt}$, as gave Eq. \eqref{rhoalfarelationship}). In this paper, the auto-regressive functions in Eq. \eqref{TheDynamicsEpModel} are of Auto-Regressive Integrated Moving Average (ARIMA) kind. The ARIMA approach is widely used for prediction of epidemic diseases, as shown by \citet{Kirbas2020}. It follows that, from a time-series viewpoint, the ARIMA model can express the evolving of a given variable (in this case, the epidemiological parameters) based on prior values. Such models, then, are coherent with the prequel discussion regarding the convergence of these parameters to steady-state conditions (Sec. \ref{modelextensionsec}).

For simplicity, instead of presenting the ARIMA fits for the three epidemiological time-series ($\beta_{opt}$, $\gamma_{opt}$ and $\rho_{opt}$), we proceed by focusing on the SARS-CoV-2 transmission factor $\beta$. We note that equivalent steps are pursued for the other parameters. The main purpose of this third layer is to model the trends of the SIRD epidemiological parameters (as provided by the two previous layers) and use these trends, in the fashion of Eq. \eqref{TheDynamicsEpModel}, to improve the forecasting of the SIRD$+$ARIMA model, making it more coherent and consistent for feedback control strategies.

It is worth mentioning that this layer is an innovative and important advantage of the SIRD model identification proposed in this work. As depicted by \citet{Villaverde2020estimating}, using a SIRD model with time-varying epidemiological parameters allows one to provide forecasts that differ at the initial and long-term moments of the COVID-19 contagion.

\subsubsection{ARIMA Fit for the Viral Transmission Rate $\beta$}

As exploited in Section \ref{sec2}, the transmission rate parameter $\beta$ gives an important measure to analyze the pandemic panorama. It has been shown that this parameter varies according to health measures applied to the prone population \cite{Villaverde2020estimating,Oliveira2020,Caccavo2020}. The ARIMA expression is given as follows:
\begin{eqnarray}\label{ARIMA_model}
    \left(a_{\beta_0} + a_{\beta_1}\beta(k-1)+a_{\beta_2}\beta(k-2) + \dots + a_{\beta_{n_\beta}}\beta(k-n_\beta + \epsilon(k)\right)\boldsymbol{\delta \beta}_{k} &=& \\ \nonumber \left(b_{\beta_0}\epsilon(k) + b_{\beta_1}\epsilon(k-1)+b_{\beta_2}\epsilon(k-2) + \dots + b_{\beta_{n_\beta}}\epsilon(k-n_\beta)\right) && \quad \text{,}
\end{eqnarray}
\noindent where $\boldsymbol{\delta \beta}_{k} \,=\ [\delta \beta(k) \; \delta \beta (k-1) \; \delta \beta (k-2) \; \dots \; \delta \beta (k-n_\beta)]^T$ is the vector of each incremental difference $\delta \beta (k) \, = \,\beta(k) - \beta(k-1)$. Notice that, regarding the total regression order of this model, in accordance with Eq. \eqref{TheDynamicsEpModel}.

Then, the ARIMA fit is performed minimizing an Extended Least-Squares (ELS) Procedure, used to find depolarized estimations for the ARIMA parameter values ($a_{\beta_0}$, $\dots$, $b_{\beta_{n_\beta}}$). This procedure is based on the following steps:
\begin{enumerate}[label =(\roman*)]
\item Concatenate the ARIMA the regression term from Eq. \eqref{ARIMA_model} as $\beta(k) = \omega^T(k-1)\Theta + \epsilon(k)$, where $\omega(k-1)$ concatenates the input values from the previous layers (in this case, the time-series $\beta_{opt}$) and $\Theta$ concatenates the ARIMA coefficients;
\item Determine a constrained Least-Squares estimation $\hat{\Theta}$, w.r.t. Eq. \eqref{ARIMA_model};
\item Compute the Least-Squares residue $\epsilon = \beta_{opt} - \omega\hat{\Theta}$ and
  use it as the initial guess for $\epsilon$ for the next iteration;
\item Iterate until convergence or some ($2$-norm wise) small residue is achieved.
\end{enumerate}


\subsection{Model Validation Results}

With respect to the detailed identification procedure, the datasets provided by Brazilian Ministry of Health are used for validation purposes, considering the interval from the first confirmed COVID-19 case in the country, dating February $26$th, until the data from June $30$th, $2020$. The Ministry of Health disclosed daily values for the cumulative number of cases and deaths. The data is sufficient to obtain the behaviors for $S$, $I$, $R$, and $D$. Note that the total population size of Brazil is used as the initial condition $N_0 \, \approx \, 210$ million.

Regarding the second layer, the optimization (normalization) weights are  presented in Table \ref{tabelOpti}. Moreover, the uncertainties bounds taken as $\underline{\delta}=0.95$ and $\overline{\delta}=1.05$.

\begin{table}[htbp]
    \caption{Optimization Weights.}
    \centering
	\begin{tabular}{c | c c c } 
        Parameter & $w_1$ & $w_2$ & $w_3$ \\ \hline
        Value & $1$ & $10$ & $2$ \\ 
    \end{tabular}
    \label{tabelOpti}
\end{table}

We proceed the model validation in a twofold: a) first we consider the first $100$ data points to tune the SIRD$+$ARIMA model and demonstrate its validity, globally well representing the following points; and b) we tune the SIRD$+$ARIMA model by identifying its parameters for the whole dataset; this is the model used for control. We note that the values used for the social isolation factor $\psi$ are those as exhibited in Figure \ref{psi_br}. 

\subsection{a) Validation}

\begin{figure}[htb]
	\centering
		\includegraphics[width=\linewidth]{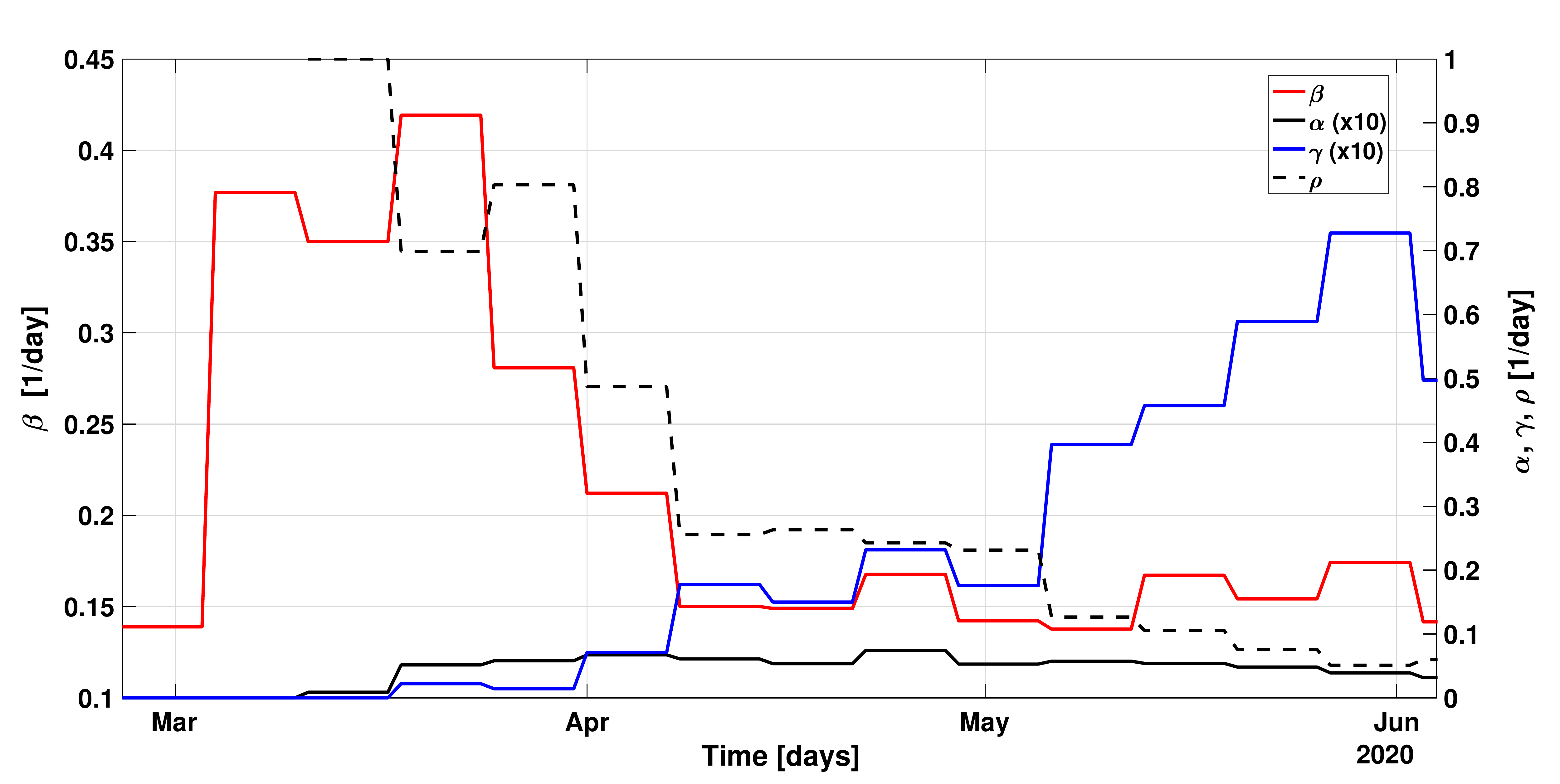}
                \caption{$\beta$, $\gamma$, $\alpha$ and $\rho$ parameters identified by the two layer approach - from February $26$th, $2020$ to June, $04$th $2020$.}
	\label{parameters}
\end{figure}

Regarding the first validation part, Figure \ref{parameters} shows the estimated time-series values from the optimization output. Recall that these parameters are piece-wise constant for periods of $T_2 \, = \, 7$ days. As discussed in previous literature, these parameters tend to follow stationary trends as the pandemic progresses. 


\begin{figure}[htb]
	\centering
		\includegraphics[width=\linewidth]{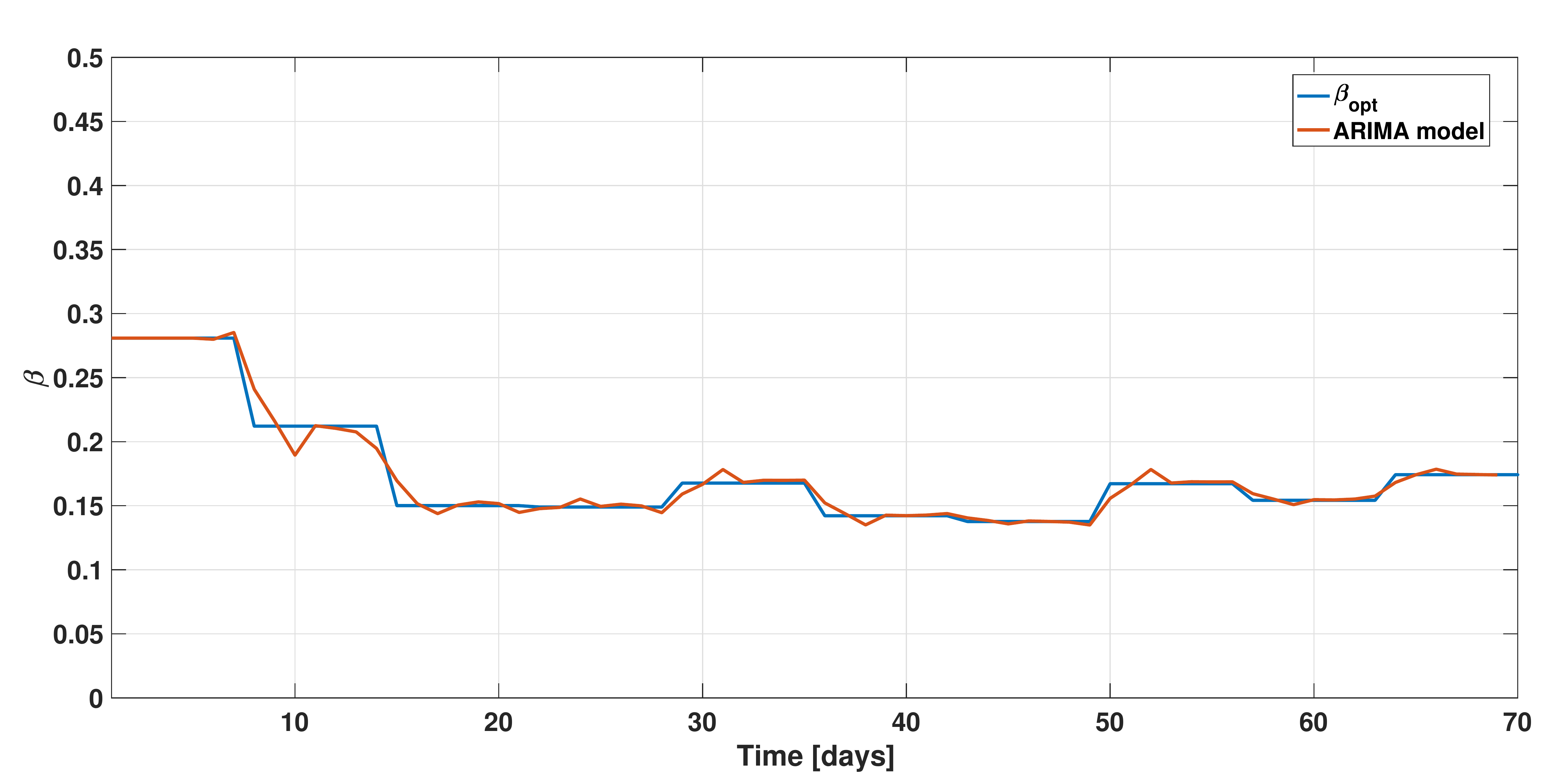}
                \caption{Comparison between ARIMA model and previous identified $\beta$. FIT - 99.04\%}
	\label{fit_beta}
\end{figure}

In order to illustrate the ARIMA fits, the identified auto-regressive model for SARS-CoV-2 transmission parameter $\beta$ is provided in Figure \ref{fit_beta}. The ELS procedure yields a regression $f_\beta(\cdot)$ with $n_\beta \, = \, 21$ daily steps (i.e. $3$ weeks). Figure \ref{fit_beta} demonstrates how the ARIMA is indeed able to describe the time-series $\beta_{opt}$, globally catching the time-varying behavior with $99.04$ \% accuracy.

\begin{figure}[htb]
	\centering
		\includegraphics[width=\linewidth]{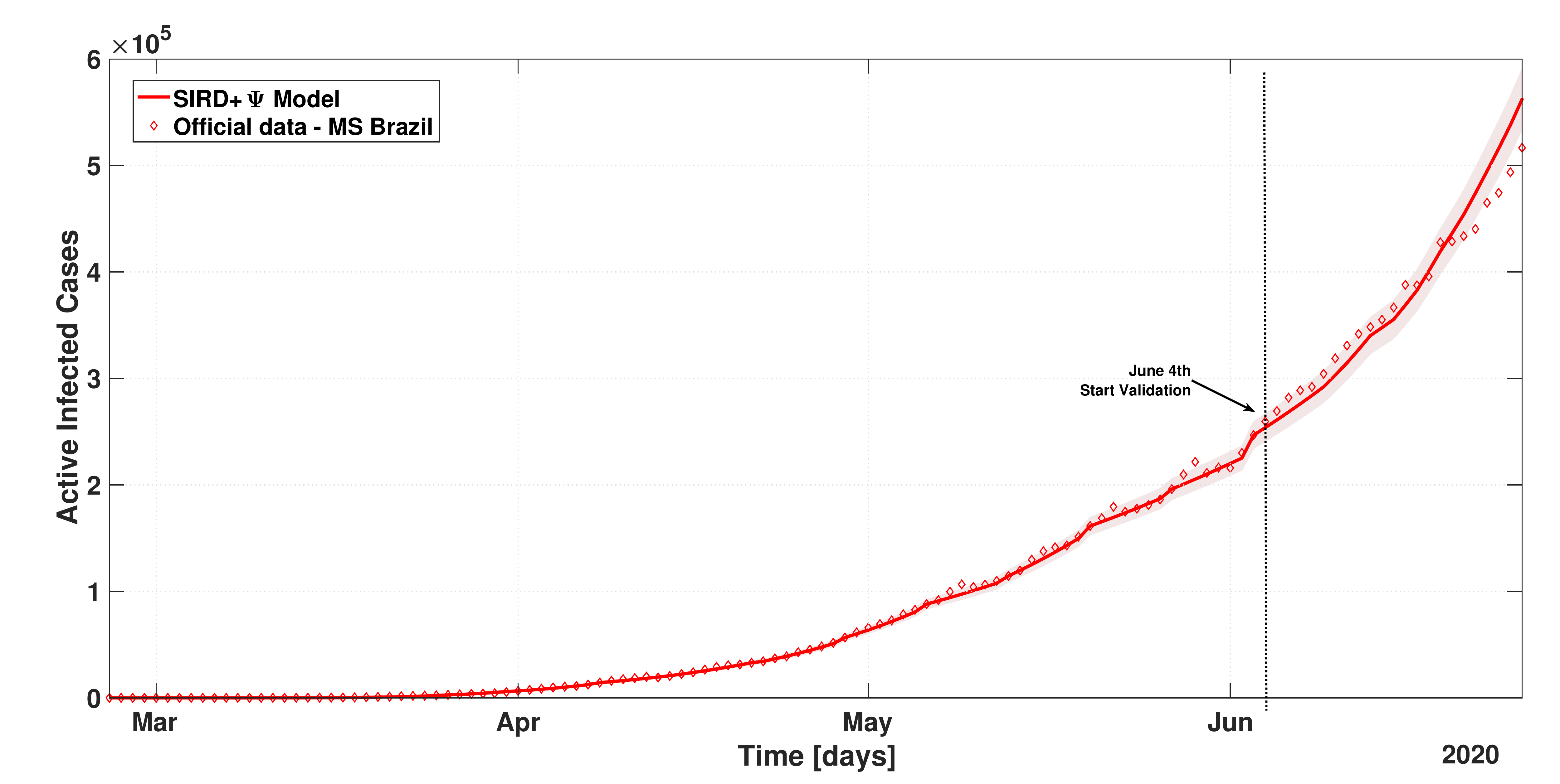}
                \caption{Validation of the SIRD$+$ARIMA model using estimated parameters with official data - Active Infected Curve.}
	\label{I_SIRD}
\end{figure}

\begin{figure}[htb]
	\centering
		\includegraphics[width=\linewidth]{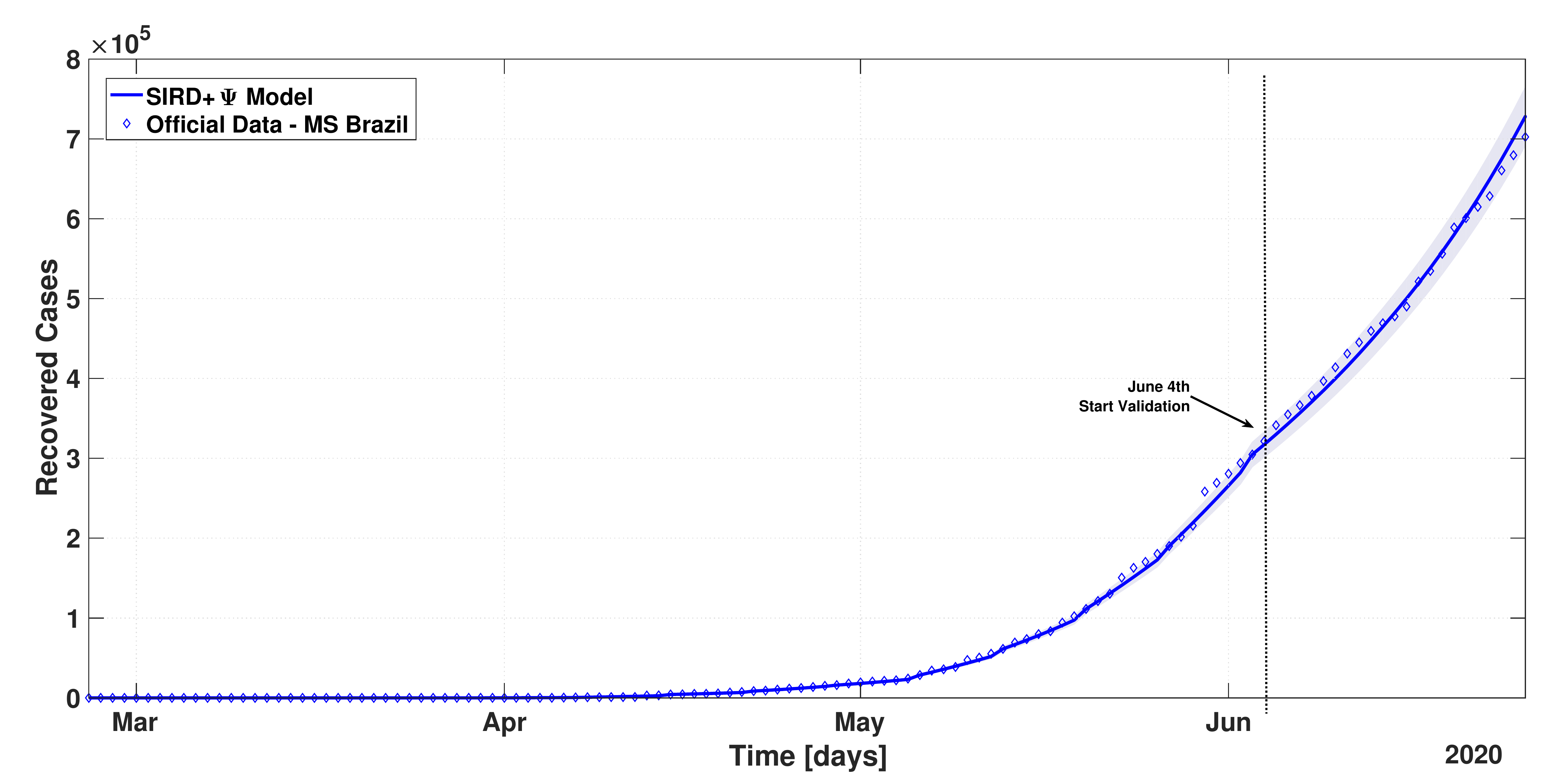}
                \caption{Validation of the SIRD$+$ARIMA model using estimated parameters with official data - Recovered Curve.}
	\label{R_SIRD}
\end{figure}

\begin{figure}[htb]
	\centering
		\includegraphics[width=\linewidth]{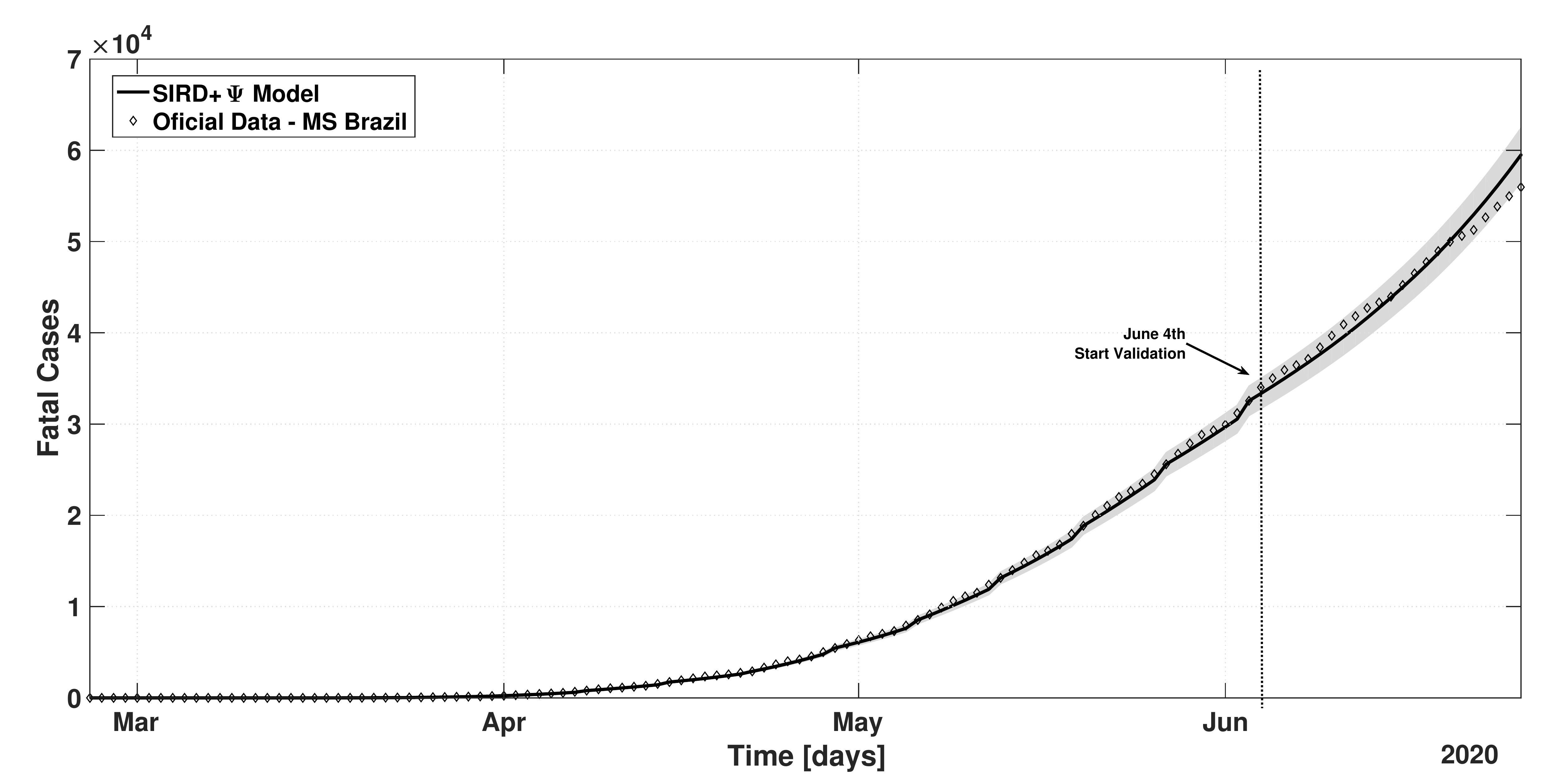}
                \caption{Validation of the SIRD$+$ARIMA model using estimated parameters with official data - Fatal Cases Curve.}
	\label{D_SIRD}
\end{figure}

\begin{figure}[htb]
	\centering
		\includegraphics[width=\linewidth]{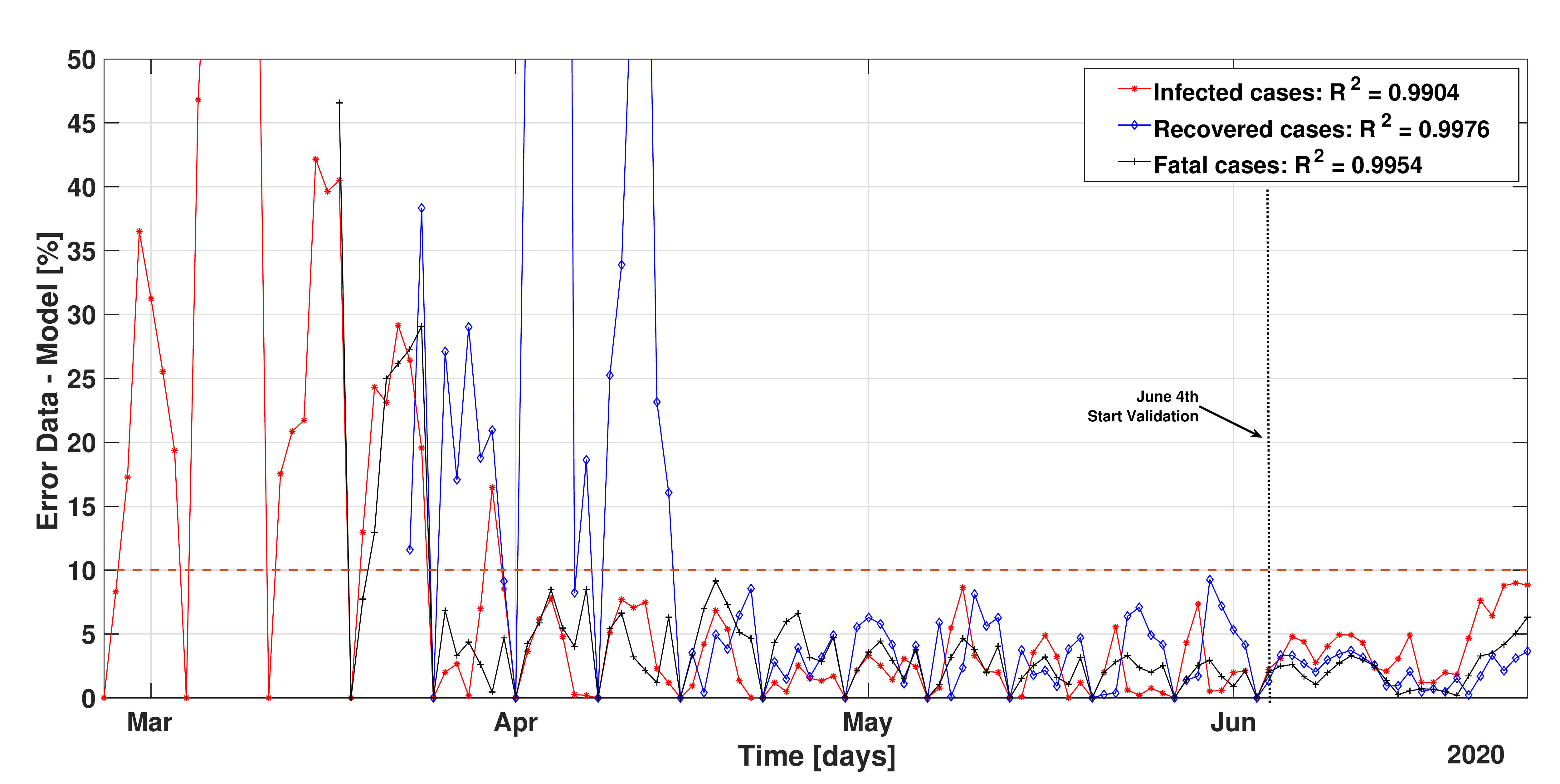}
                \caption{Error between the simulated model and official data (in percentage).}
	\label{error}
\end{figure}

The first $100$ data points comprise the period from February $26$th to June, $04$th $2020$. Using the resulting SIRD$+$ARIMA model identified in this period, we proceed by extending the model forward, making forecasts from June, $05$th until June $30$th, $2020$, in order to demonstrate the validity of the identification. The forecasts are made using Eqs. \eqref{eqSIRDmodel} wherein parameters $\gamma$, $\beta$ and $\rho$ are given by the ARIMA regressions model in Eq. \eqref{TheDynamicsEpModel}. Figures \ref{I_SIRD}, \ref{R_SIRD} and \ref{D_SIRD} show the model forecasts compared with real data for active infections, recovered individuals and deaths, respectively, considering a interval of confidence of $95$\%. Clearly, the global behavior of the COVID-19 contagion is well described by the proposed models. To further illustrate this fact, Figure \ref{error} depicts the model-data error terms (from Eqs. \eqref{eq:err}-\eqref{eq:err3}), given in percentage. The coefficient of determination for these identification results are all above $0.99$.

\subsection{b) Data-driven SIRD$+$ARIMA Model}

\begin{figure}[htb]
	\centering
		\includegraphics[width=\linewidth]{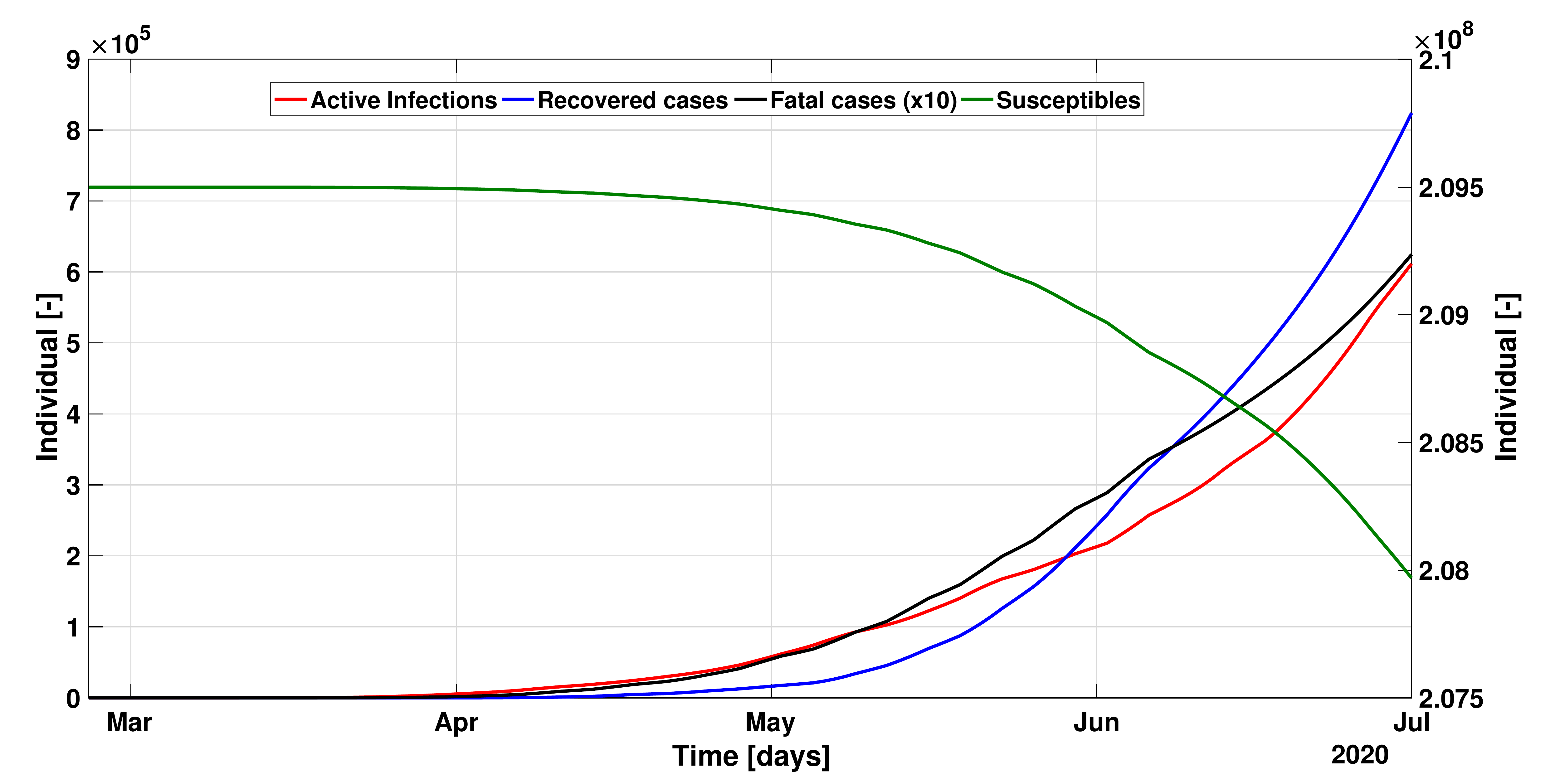}
                \caption{SIRD model curves used for NMPC framework. Right axis for Susceptible individuals (S). Left axis for Active Infections (I), Recovered (R) and Fatal (D) Cases.}
	\label{Model_SIRD}
\end{figure}

We note that the data-driven model, that is used for the NMPC strategy, is the based on the whole set of data. The simulation of this model is given in Figure \ref{Model_SIRD}. 

Based on the prior developments, given the high coefficient of determination and observing Figures \ref{fit_beta}-\ref{error} (in which a $95$\% confidence interval is considered), it can be noticed that the identified model SIRD$+$ARIMA describes very well the observed COVID-19 pandemic behavior in Brazil and, in addition, it can forecast the SIRD curve with relatively small error. It is worth mentioning that this model identification approach needs to be updated regularly. From Figure \ref{error}, one notes that $20$-days-ahead forecasts can be derived with model-data errors below $10$\%. This error margins may increase when considering larger forecast horizons. Nevertheless, regarding the control purpose of this work and taking into account prediction horizons of roughly one month, the proposed SIRD$+$ARIMA is indeed able to be highly representative regarding the COVID-19 spread. For perfectly coherent forecasts, it is recommendable for the identification to be re-performed each week, and the ARIMA fits updated. This recommendation has also been provided by \citet{morato2020optimal}.

We also remark that the ARIMA fits for the epidemiological parameters are given in weekly samples i.e. $k_{\text{week}}$, while the SIRD variables are given for each day i.e. $k_{\text{day}}$. Regarding this matter, we can represent the weekly-sampled variables, from the viewpoint of the daily-sampled variables, as:
\begin{eqnarray} \nonumber
\beta\left(k_{\text{day}}\right) &=& \beta\left(k_{\text{week}}T_2\right) \, \, \, \forall \, k_{\text{day}} \, \in \, \left[k_{\text{week}}T_2 \, , \, (k_{\text{week}}+1)T_2\right) \quad \text{,}\\ \ \nonumber
\gamma\left(k_{\text{day}}\right) &=& \gamma\left(k_{\text{week}}T_2\right) \, \, \, \forall \, k_{\text{day}} \, \in \, \left[k_{\text{week}}T_2 \, , \, (k_{\text{week}}+1)T_2\right) \quad \text{,} \\
\rho\left(k_{\text{day}}\right) &=& \rho\left(k_{\text{week}}T_2\right) \, \, \, \forall \, k_{\text{day}} \, \in \, \left[k_{\text{week}}T_2 \, , \, (k_{\text{week}}+1)T_2\right) \quad \text{,} \\ \nonumber
\psi\left(k_{\text{day}}\right) &=& \psi\left(k_{\text{week}}T_2\right) \, \, \, \forall \, k_{\text{day}} \, \in \, \left[k_{\text{week}}T_2 \, , \, (k_{\text{week}}+1)T_2\right) \quad \text{,} \\ \nonumber
u\left(k_{\text{day}}\right) &=& u\left(k_{\text{week}}T_2\right) \, \, \, \forall \, k_{\text{day}} \, \in \, \left[k_{\text{week}}T_2 \, , \, (k_{\text{week}}+1)T_2\right) \quad \text{,}
\end{eqnarray}
where $T_2 \, = \, 7$ days.

\begin{figure}[htb]
	\centering
		\includegraphics[width=\linewidth]{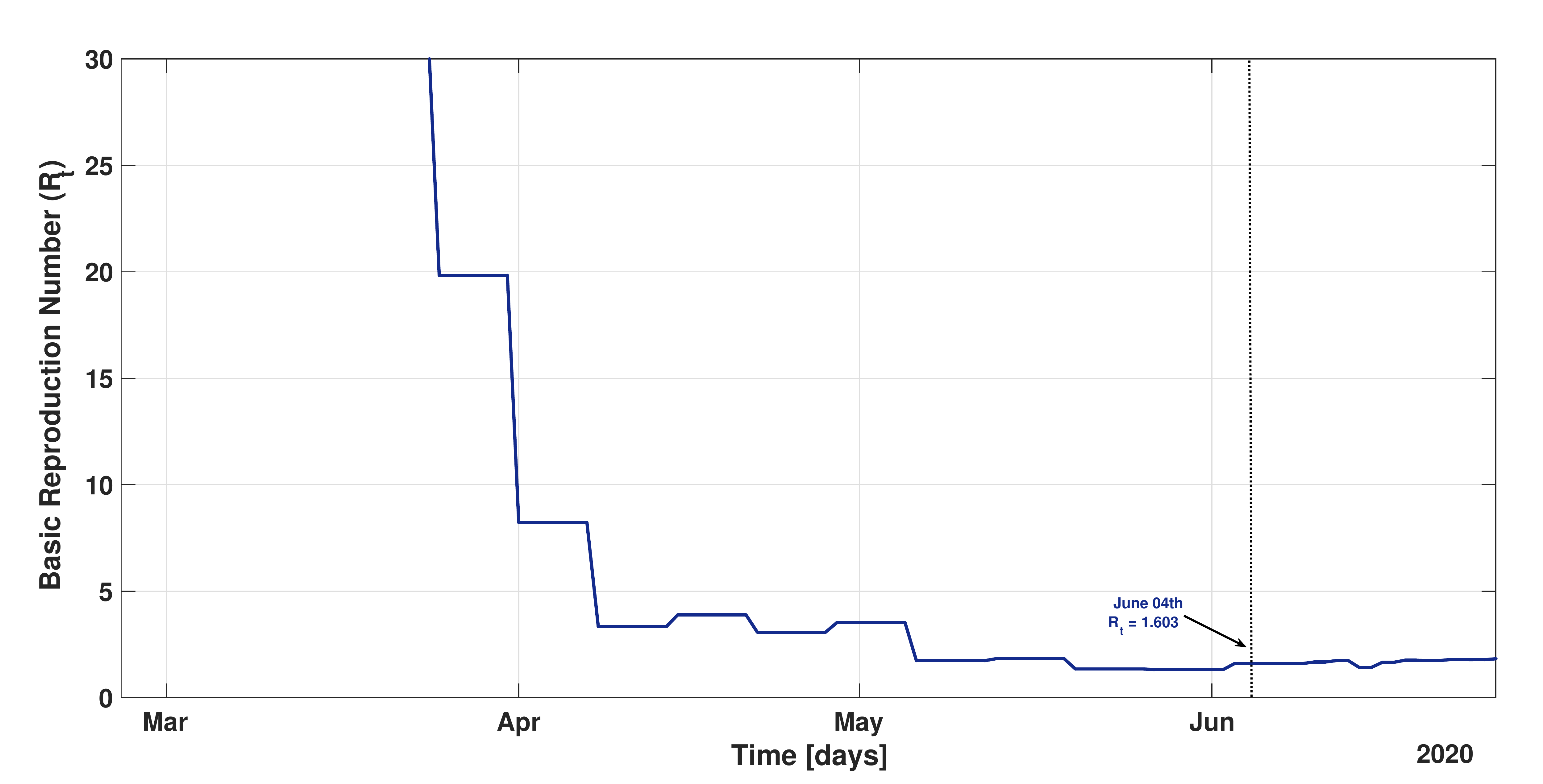}
                \caption{Basic reproduction number $R_t$ calculated by the multi-layered identification framework.}
	\label{r0}
\end{figure}

Finally, considering this validated SIRD$+$ARIMA model, the basic reproduction number $R_t$ of the SARS-CoV-2 virus in Brazil can be inferred through Eq. \eqref{R0eq}. In Figure \ref{r0}, the evolution of the viral reproduction number $R_t$ for each week of the pandemic in show.  As it can be seen, this reproduction number presents stronger variations at the beginning stage of, but then tends to converge to a steadier values, which corroborates with the model extensions considering the epidemiological parameters. Moreover, notice that the reproduction number for since June $4$th, $2020$ is somewhat steady near $1.603$, which indicates that the COVID-19 pandemic is still spreading in the country. 

\FloatBarrier
\section{The NMPC strategy}
\label{sec3}

In this Section, we detail the proposed NMPC strategy used to determine public health guidelines (regarding social isolation policies) to mitigate the spread of the COVID-19 contagion in Brazil. This NMPC algorithm is conceived under the feedback structure illustrated in Figure \ref{modeltotal}.

We note that the generated control action $u(k)$ represents the input to the population's response to social distancing measures, as gives the differential Eq. \eqref{TheControlPsi}. For this reason, it follows that the proposed NMPC algorithm operates with a weekly sampling period ($T_2$). Implementing a new social distancing guideline every week is coherent with previous discussions in the literature \cite{morato2020optimal}. It does not seem reasonable to change the distancing measures every few days.

\subsection{Possible Control Sequences}

Before presenting the actual implementation of NMPC tool, we note that its generated control sequence must be given in accordance with the constraints expressed in Eq. \eqref{UConstraints}.

Considering that the NMPC has a horizon of $N_p$ steps (given in weekly samples) from the viewpoint of each week $k$, the generated control signal is\footnote{Notation $\chi(k+j|k)$ denotes the predicted values for variable $\chi(k+j)$, computed at the discrete instant $k$.}:

\begin{eqnarray}
\label{BigUSequence}
U_{k} &=& \left[\begin{array}{cccc} u(k|k) & u(k+1|k) & \cdots & u(k+N_p-1|k)\end{array}\right] \quad \text{.}
\end{eqnarray}

Since the variations from each quarantine guideline $u (k-1)$ to the following $u(k)$, denoted $\delta u(k)$ are equal to $\pm \, 0.05$ or $0$, it follows that all possible control sequences can be described as, from the viewpoint of sample $k$:
\begin{eqnarray}
\label{BigUSequence2}
U_{k} &=& \left[\begin{array}{ccc} \left(u(k-1) + \delta u(k)\right) & \cdots & \left(u(k-1) \overbrace{+ \delta u(k) \dots + \delta u(k+N_p-1)}^{N_p\text{ times}}\right) \end{array}\right] \, \text{.}
\end{eqnarray}

From Eq. \eqref{BigUSequence2}, we can conclude that, in the considered settling, there are $3^{N_p}$ possible control sequences.

\subsection{Control Objectives}

The main purpose of social isolation is to distribute the demands for hospital bed over time, so that all infected individuals can be treated. It seems reasonable to act though social isolation to minimize the number of active infections ($I$), while ensuring that this class of individuals remains smaller than the total number of available ICU beds $n_{ICU}$.

Moreover, it seems reasonable to ensure that social isolation measures are enacted for as little time as possible, to mitigate the inherent economic backlashes.

This trade-off objective (mitigating $I$ against relaxing social isolation) is expressed mathematically as:
\begin{eqnarray}
\label{TheCO}
\left|\left|I(k)\right|\right|_{\frac{Q}{n_I^2}} + \left|\left|u(k)\right|\right|_{(1-Q)} \,\,=\,\, I(k)^T\frac{Q}{n_I^2}I(k) + u(k)^T(1-Q)u(k) \quad \text{,}
\end{eqnarray}
where $n_I$ is a nominal limit for $I$ (i.e. the initial population size $N_0$) included for a magnitude normalization of the trade-off objective. Note that $u$ is given within $[0.3 \, \, 0.7]$, so there is no need for normalization.
\subsection{Process Constraints}

The proposed NMPC algorithm must act to address the Control Objective given in Eq. \eqref{TheCO} while ensuring some inherent process constraints. These are:
\begin{enumerate}
    \item Ensure that the control signal has the piece-wise constant characteristic, as gives Eq. \eqref{UConstraints}; and
    \item Ensuring that the number of infected people with active acute symptoms does not surpass the total number of available ICU hospital beds in Brazil, this is:
    \begin{eqnarray}
    \label{constInICU}
    p_{\text{sym}}I(k) &\leq& n_{ICU} \text{.}
    \end{eqnarray}
\end{enumerate}

\subsection{The Complete NMPC Optimization}   

Bearing in mind the previous discussions, the NMPC procedure is formalized under the following optimization problem, which is set to mitigate the SARS-CoV-2 viral contagion spread with respect to the trade-off control objective of Eq. \eqref{TheCO} and the constraints from Eqs. \eqref{UConstraints}-\eqref{constInICU}, for a control horizon of $N_p$ (weekly) steps, from the viewpoint of each sampling instant $k$:
\begin{eqnarray}
\label{TheCompleteNMPC}
    \underset{U_{k}}{\min}  \, J(\cdot) &=& \underset{U_{k}}{\min} \sum_{i=1}^{N_p} \left( \left(I(k+i)^T\frac{Q}{n_I^2}I_j(k+i)\right) + \left(u(k+i-1)^T(1-Q)u(k+i-1)\right)\right) \text{,} \\ \nonumber
    \text{s.t.:} && \text{SIRD$+$ARIMA Model} \,\,\, \forall \, i \in \, \mathbb{N}_{[1\, , \, N_p]} \,\text{,} \\ \nonumber
    && \underline{\psi} \leq u(k+i-1) \leq \overline{\psi} \,\text{,}\\ \nonumber
    && u(k+i-1) \, = \, u(k+i-2) + \delta u(k+i-1) \, \text{,} \\ \nonumber
    &&  \delta u(k+i-1) \, = \, -0.05\text{ or }0\text{ or }0.05 \,\text{,}\\ \nonumber
&& p_{\text{sym}}I(k+i)  \leq n_{ICU} \,\text{.}
\end{eqnarray}

\subsection{Finitely Parametrized NMPC Algorithm}

The finitely parametrized NMPC methodology has been elaborated by \citet{alamir2006stabilization}. This paradigm has recently been extended to multiple applications \cite{rathai2018parameterized,rathai2019gpu}, with successive real-time results.

Given that this control framework offers the tool to formulate (finitely parametrized) social distancing guidelines for the COVID-19 spread in Brazil, with regard to the discussions in the prior, we proceed by detailing how it is implemented at each sampling instant $k$, ensuring that the Nonlinear Optimization Problem in Eq. \eqref{TheCompleteNMPC} is solved.

\begin{figure}[htb]
	\centering
		\includegraphics[width=\linewidth]{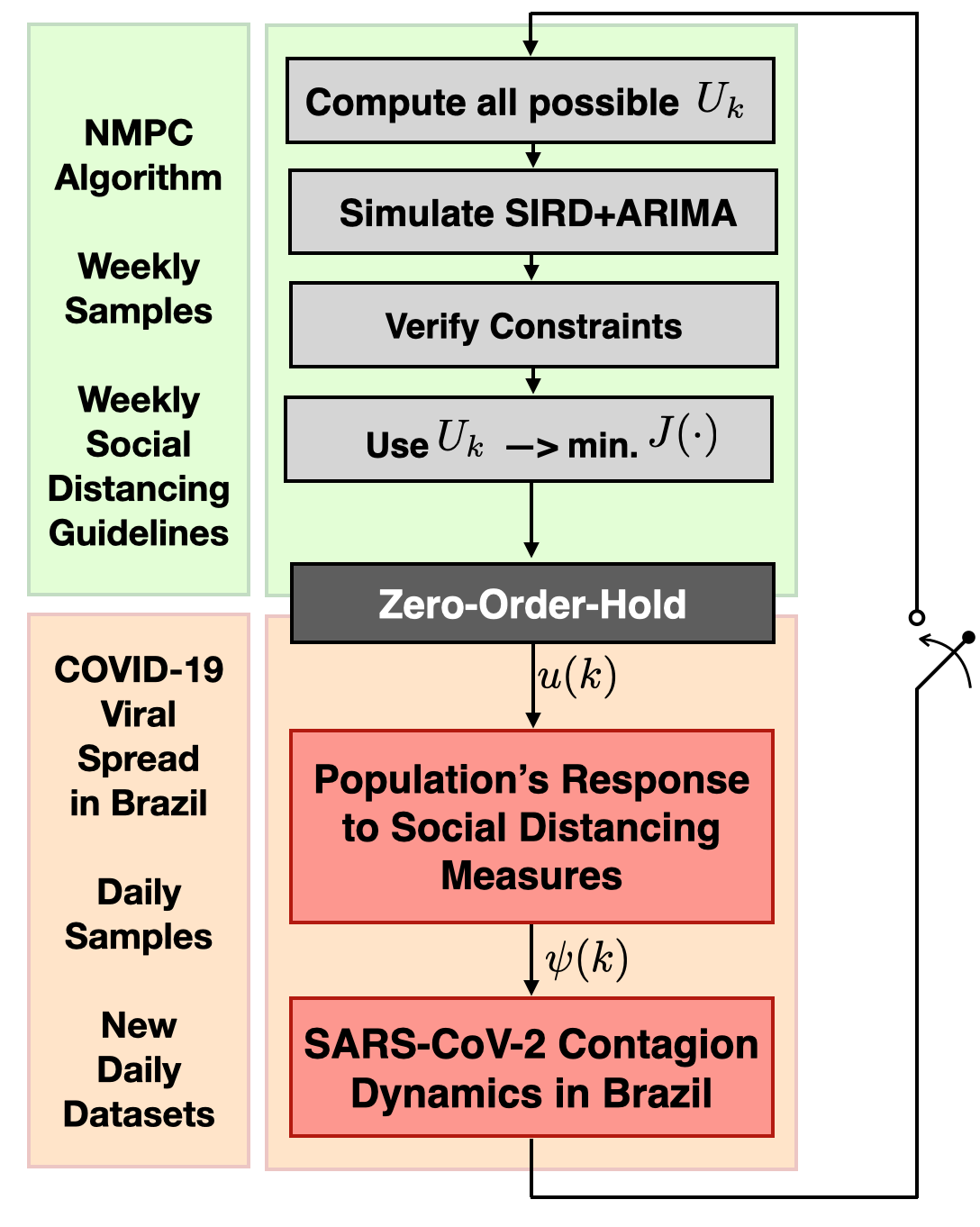}
                \caption{Algorithm Implementation Flowchart.}
	\label{ThepNMPCAlgoCovid}
\end{figure}

Basically, the parametrized NMPC algorithm is implemented by simulating the SIRD$+$ARIMA model with an explicit nonlinear solver, testing it according to all possible control sequences (as gives Eq. \eqref{BigUSequence2}). Then, the predicted variables are used to evaluate the cost function $J(\cdot)$. The control sequences that imply in the violation of constraints are neglected. Then, the resulting control sequence is the one that yields the minimal $J(\cdot)$, while abiding to the constraints. Finally, the first control signal is applied and the horizon slides forward. This paradigm is explained in the Algorithm below. Figure \ref{ThepNMPCAlgoCovid} illustrates the flow of the implementation of the proposed Social Distancing control methodology for the COVID-19 viral spread in Brazil.

\begin{algorithm}{\textbf{Finitely Parametrized NMPC for Social Distancing Guidelines}} \label{ThepNMPCAlgoCovidalgo}\\  \noindent\makebox[\linewidth]{\rule{\linewidth}{0.4pt}}
\textbf{Initialize}: $N(0)$, $S(0)$, $I(0)$, $R(0)$ and $D(0)$. \\
\textbf{Require}: $Q$, $n_I$, $n_{ICU}$ \\
\textbf{Loop every $T_1$ days}:
\begin{itemize}
\item Step (\textit{i}): "Measure" the available contagion data ($N(k)$, $S(k)$, $I(k)$, $R(k)$ and $D(k)$);
\item Step (\textit{ii}): \textbf{Loop every $T_2$ days}:
\begin{itemize}
\item Step (a): \textbf{For each sequence $j \, \in \, 3^{N_p}$}:
\begin{itemize}
    \item Step (1): Build the control sequence vector according to Eq. \eqref{BigUSequence2};
    \item Step (2): Explicitly simulate the future sequence of the SIRD variables;
    \item Step (3): Evaluate if constraints are respected. If they are not, end, else, compute the cost function $J(\cdot)$ value.
\end{itemize}
\item \textbf{end}
\item Step (b): Choose the control sequence $U_{k}$ that corresponds to the smallest $J(\cdot)$.
\item Step (c): Increment $k$, i.e. $k \leftarrow k+1$.
\end{itemize}
\item \textbf{end}
\item Step (\textit{iii}): Apply the local control policy $u(k)$ as gives Eq. \eqref{UConstraints};
\item Step (\textit{iv}): Simulate the SIRD$+$ARIMA model, as give Eqs. \eqref{eqSIRDmodel}-\eqref{TheControlPsi}.
\item Step (\textit{v}):  Increment $k$, i.e. $k \leftarrow k+1$.
\end{itemize}
\textbf{end}
\end{algorithm}

 \FloatBarrier

\section{Simulation Results}
\label{sec4}

Bearing in mind the previous discussions, the proposed SIRD$+$ARIMA model and the finitely parametrized NMPC toolkit, this Section is devoted in presenting the simulation results regarding the COVID-19 contagion spread in Brazil. The following results were obtained using Matlab. The average computation time needed to solve the solution of the proposed NMPC procedure is of $2$ ms\footnote{The implementation was performed on a i5 $2.4$ GHz Macintosh computer.}.

\subsection{Simulation Scenarios}

We proceed by depicting two simulation scenarios, which account for the following situations:
 \begin{enumerate}[label=(\alph*)]
    \item What would be the case if the NMPC generated social distancing policy was enacted back in the beginning of the COVID-19 pandemic in Brazil.
    \item How plausible is it to apply the NMPC technique from now on ($30/06/2020$) and reduce and mitigate the effected of the spread.
\end{enumerate}

In order to provide these scenarios, we consider an ``open-loop" comparison condition, which represents the simulation of the SIRD$+$ARIMA model with the social distancing factor in fact observed in Brazil (see Figure \ref{psi_br}). The results are based on $126$ days of data. From the $127^{th}$ sample on, the open-loop simulation takes $\psi = 0.3$ (no isolation).

The proposed NMPC is designed with a cost function $J(\cdot)$ with tuning weight $Q \, = \, 0.7$ (which means reducing the number of infections is prioritized over relaxing social distancing). Furthermore, the NMPC optimization is set with a prediction horizon of $N_p \, = \, 4$ weeks ($28$ days), which means that the controller makes its decision according to model-based forecasts of the SIRD$+$ARIMA model for roughly one month ahead of each sample $k$. For simplicity $n_I$ is taken as  $\frac{n_{ICU}}{p_{\text{sym}}}$, since the main goal of the MPC is to ensure $p_{\text{sym}}I \, \leq \, n_{ICU}$.

\subsection{Scenario (a): NMPC since the beginning}

We recall that the COVID-19 pandemic in Brazil formally ``started" $26/02/2020$, when the first case was officially registered.

This first simulation scenario considers the application of the NMPC control strategy to guide social distancing from $30$ days after the first case ($27/03/2020$). We choose this date because, in Brazil, the majority of states started social distancing measures (in different levels) around the period of late-March/mid-April. We do not deem it reasonable to consider the application of the NMPC strategy from ``day $1$", since this was not seen anywhere in Brazil.

\begin{figure}[htb]
	\centering
		\includegraphics[width=\linewidth]{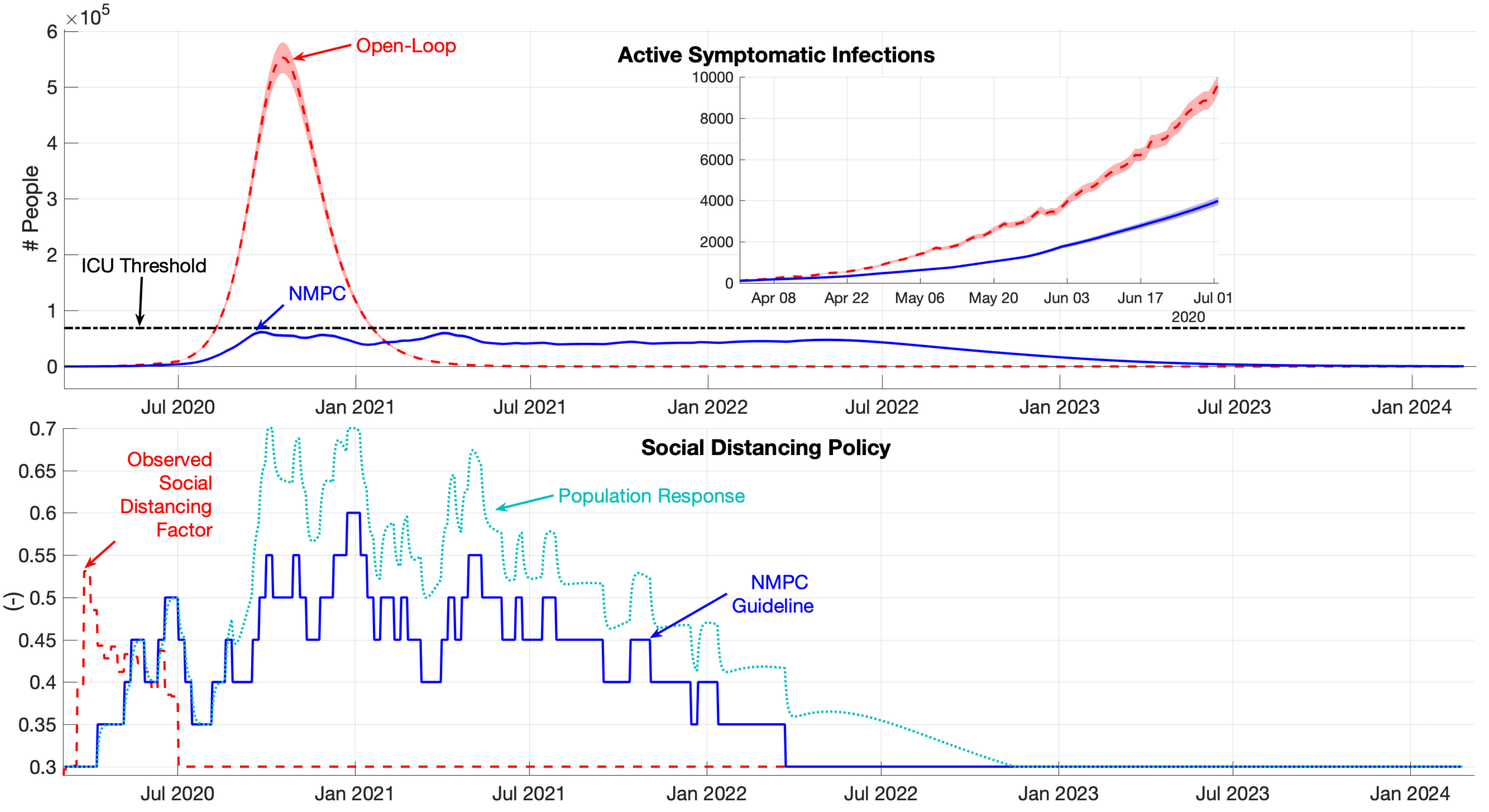}
                \caption{Scenario (a): Symptomatic Active Infections $p_{\text{sym}}I$ and Control Input $u$ (Social Distancing, $\psi$).}
	\label{WhatIfActiveU}
\end{figure}

Considering this control paradigm, Figure \ref{WhatIfActiveU} depicts the predicted evolution of severe/acute symptomatic COVID-19 infections over time. These active symptomatic infections stand for those that may need ICU hospitalization. The results shows that the NMPC is able to thorough attenuate the peak of infections, ensuring that it always stays below the ICU threshold. This is a quite significant results, which shows that the proposed feedback framework is able to offer an enhanced paradigm, with time-varying social distancing measures, such that the COVID-19 spread curve is indeed flattened, never posing serious catastrophic difficulties to Brazilian hospitals (and health system overall). The result also indicated that the social distancing should vary with relaxing and strengthening periods over the years of $2020$, $2021$ and $2022$. We note, once again, that health professionals should better qualify the relationship between the social distancing factor and actual economic/social restrictions, as illustrated in Table \ref{TheTableUParametre}. The results also indicate that if no coordinated action is deployed by the Brazilian federal government, the number of active symptomatic infections at a given day could be up to $540000$ individuals, with this peak forecast to October $16$th, $2020$. If the NMPC strategy was indeed applied, two peaks would have been seen, with $80\, \rm{\%}$ of total ICU capacity, previewed for September $25$th, $2020$ and March, $31$rst, $2021$.

\begin{figure}[htb]
	\centering
		\includegraphics[width=\linewidth]{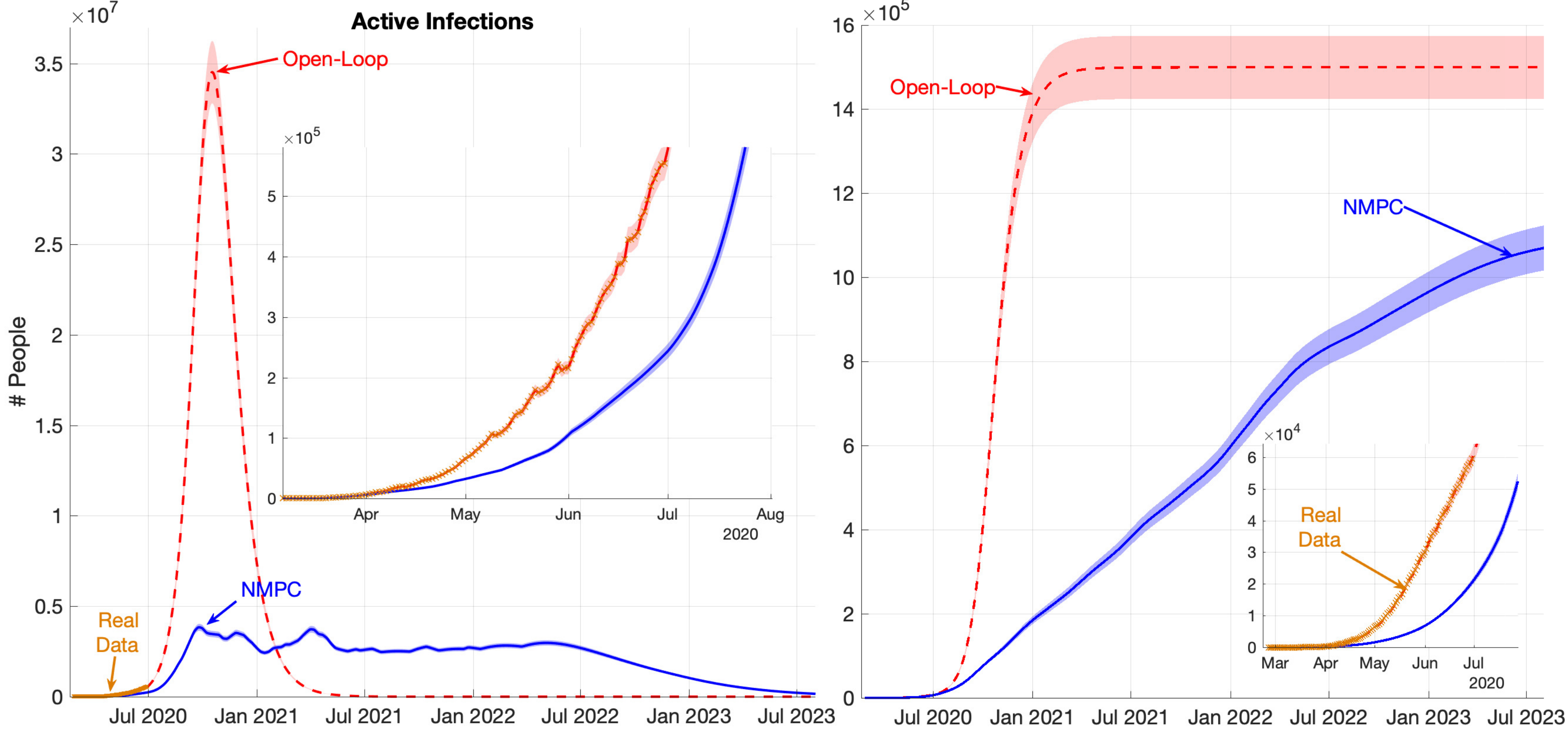}
                \caption{Scenario (a): Total Active Infections $I$ and Total Deaths $D$.}
	\label{WhatIfAllIAllD}
\end{figure}

Regarding this scenario, Figure \ref{WhatIfAllIAllD} shows the evolution of the total amount of infections and cumulative number of deaths. This Figure also places the real data points against the simulated model. The possibilities to come are catastrophic, as also previewed by \citet{morato2020optimal}. We note, once again, that the SIRD$+$ARIMA models offers qualitative results, the magnitude of $1.5$ million deaths seems quite alarming, but the model is identified considering a large margin for sub-reports (in number of deaths and confirmed cases). Not all deaths due to COVID-19 are currently being accounted for in Brazil, as discuss \citet{THELANCET20201461,zacchi:hal-02881690}.

\FloatBarrier

\subsection{Scenario (b): NMPC from the now on}

The second simulation scenario that we consider is to control the situation from the current moment ($30/06/2020$) to avoid a total collapse of the Brazilian health system. We note that as of this date, the country counts over $1.4$ million confirmed COVID-19 cases and more than $59500$ deaths due to the SARS-CoV-2 virus.

\begin{figure}[htb]
	\centering
		\includegraphics[width=\linewidth]{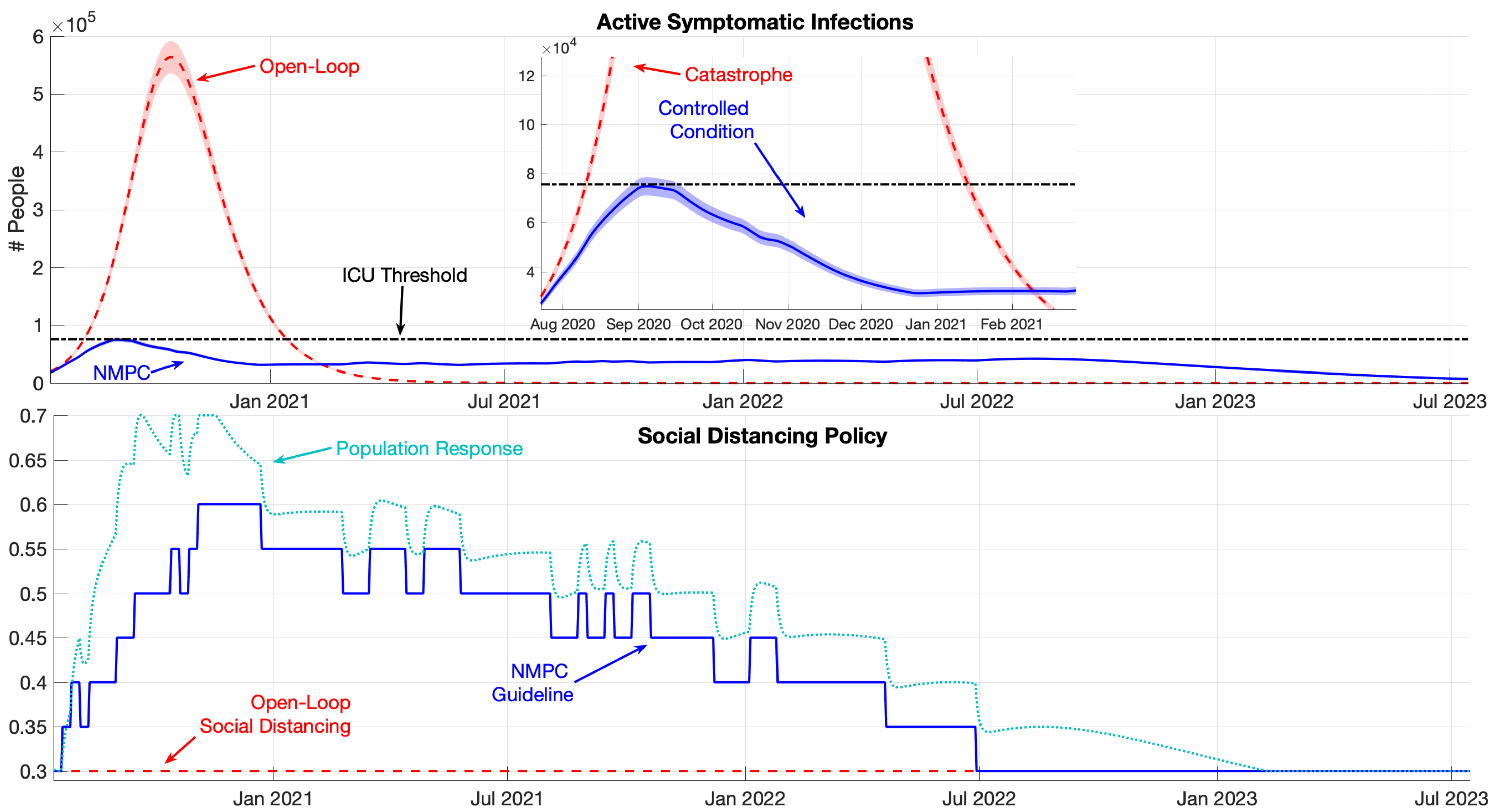}
                \caption{Scenario (b): Symptomatic Active Infections $p_{\text{sym}}I$ and Control Input $u$ (Social Distancing, $\psi$).}
	\label{FromNowActiveInfControl}
\end{figure}

\begin{figure}[htb]
	\centering
		\includegraphics[width=\linewidth]{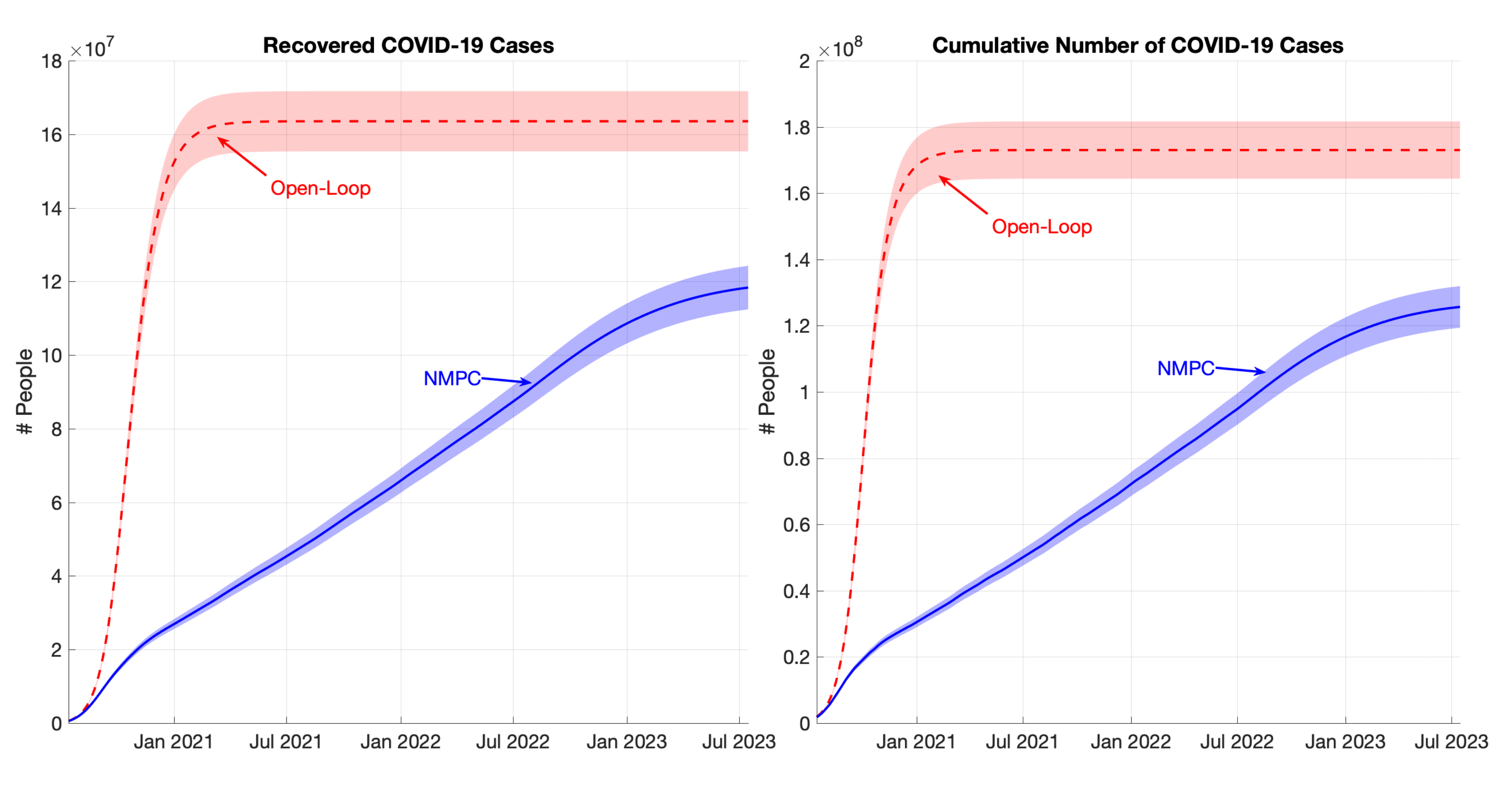}
                \caption{Scenario (b): Cumulative Cases $I_c$ and Recovered Individuals $R$.}
	\label{FromNowIcumvsR}
\end{figure}

\begin{figure}[htb]
	\centering
		\includegraphics[width=\linewidth]{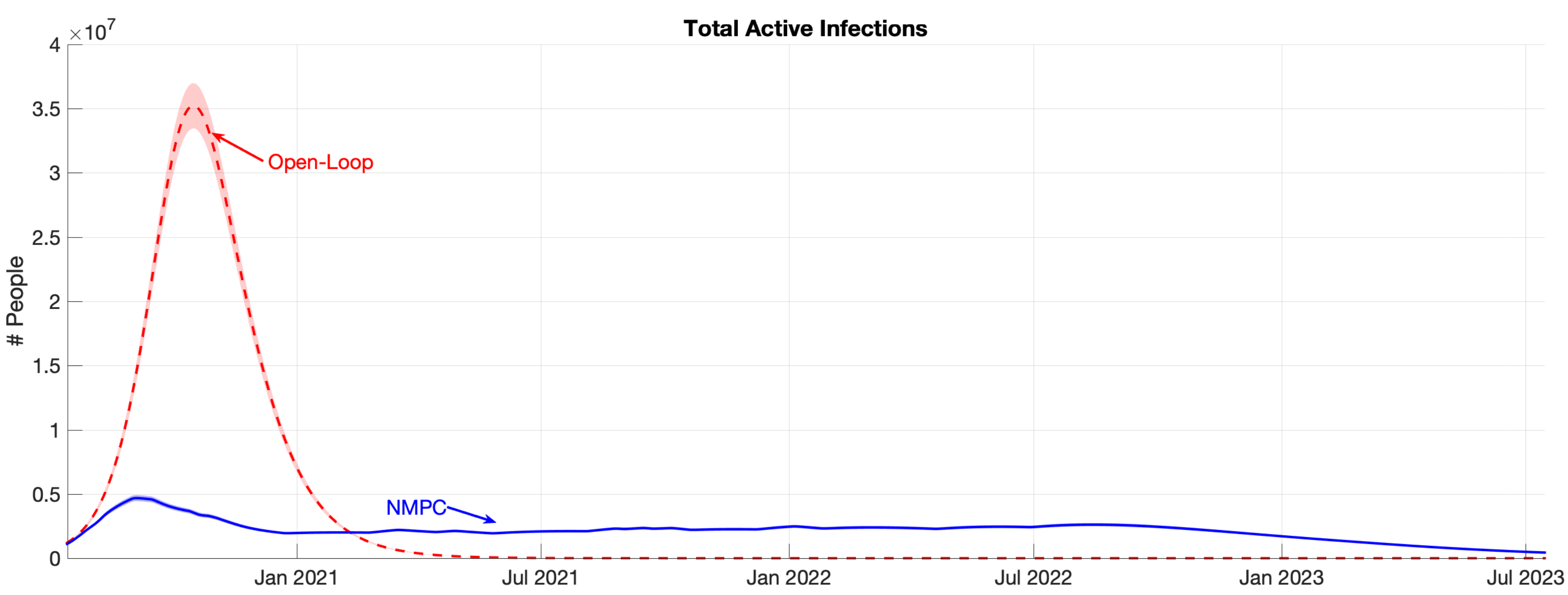}
                \caption{Scenario (b): Total Active Cases $I$.}
	\label{FromNowActiveI}
\end{figure}

\begin{figure}[htb]
	\centering
		\includegraphics[width=\linewidth]{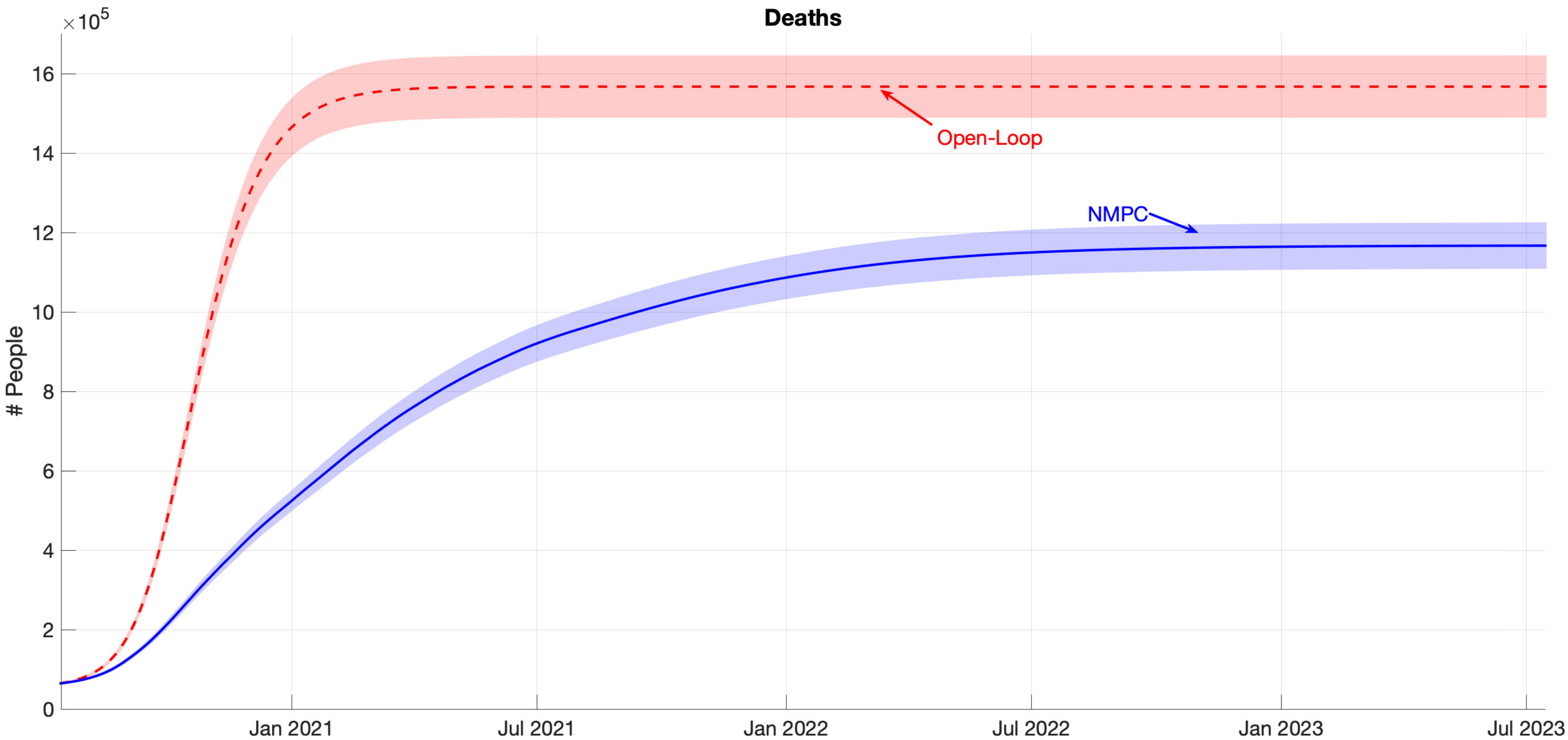}
                \caption{Scenario (b): Total Deaths $D$.}
	\label{FromNowDeaths}
\end{figure}

Figure \ref{FromNowActiveInfControl} shows the main results regarding this second scenario, considering the active symptomatic infections, that may require ICU hospitalization. Even though a partial catastrophe is already under course in Brazil, if social distancing are guided through the proposed NMPC, a total collapse of the public health system can still be avoided. Figure \ref{FromNowIcumvsR} shows the evolution for recovered individuals and cumulative number of cases $I_c$, Figure \ref{FromNowActiveI} shows the evolution of all active infections (symptomatic and asymptomatic), while Figure \ref{FromNowDeaths} shows the mounting number of deaths. The proposed NMPC, if rapidly put in practice, could still be able to slow the viral spread, saving $25 \, \rm{\%}$ of lives w.r.t. the open-loop condition. The peak of infections, if such technique is applied, has its forecast previewed to September $2$nd, 2020, being anticipated in $17$ days.

\FloatBarrier

\section{Conclusions}
\label{sec5}

In this paper, an optimal control procedure is proposed for ruling social isolation guidelines in Brazil, in order to mitigate the spread of the SARS-CoV-2 virus. To do so, a new model is proposed, based on extensions of the SIRD equations. The proposed model embeds weekly auto-regressive dynamics for the epidemiological parameters and also takes a dynamic social distancing factor (as seen in previous papers \citep{morato2020optimal,bastos2020covid19}). The social distancing factor measures and expresses the population's response to quarantine measures, guided by the Nonlinear Model Predictive Control procedure. The NMPC strategy is designed within a finitely parametrized input paradigm, which enables its fast realization.

In this work, some key insights were given regarding the future panoramas for the COVID-19 pandemic in Brazil. Below, some of the main findings of this paper are summarized:
\begin{itemize}
    \item The presented results corroborate the hypothesis formulated in many of the previous papers regarding the COVID-19 pandemic in Brazil \cite{hellewell2020feasibility, THELANCET20201461, morato2020optimal}: herd immunity is not a plausible option for the country\footnote{Previous paper have also elaborated on the fact that vertical isolation is also not an option for the time being, since Brazil does not have the means to pull of efficient public policies to separate the population at risk from those with reduced risk, due to multiple social-economical issues of the country \cite{silva2020bayesian,rocha2020expected,rodriguez2020covid}.}; if no coordinates social distancing action is enforced, the ICU threshold will be largely surpassed, which can lead to elevated fatality. 
    \item The prediction of evolution of the viral spread is relatively accurate with the proposed adapted SIRD model for up to $20$ days. Larger prediction horizons can be considered, but daily model-updates are recommended.
    \item The simulation forecasts derived with the NMPC strategy and with an open-loop condition (no social distancing) indicate that social distancing measures should still be maintained for a long time. The strength of these measures will be diluted as time progresses. The forecasts indicate an infection peak of over $600000$ symptomatic individuals to late September, $2020$, in the current setting. If model-based control is enacted, the peak could be anticipated and the level of infections could be contained below the ICU hospital bed threshold. The NMPC could save over $400000$ lives if enacted from now (July, $2020$).
    \item The results also indicate that if such coordinated control strategy was applied since the first month of COVID-19 infections in Brazil, a more relaxed social distancing paradigm would be possible as of late $2020$. Since this has not been pursued, the social distancing measures may go up until late $2021$ if no vaccine is made available.
\end{itemize}

These results presented in this paper are qualitative. Brazil has not been testing enough its population (neither via mass testing or sampled testing), which means that the data regarding the number of infections is very inconsistent. As \citet{bastos2020covid19} thoroughly details, the uncertainty margin associated to the available data (in terms of case sub-reporting) is very significant. Anyhow, the results presented herein can help guiding long-term regulatory decision
policies in Brazil regarding COVID-19. One must note that social distancing measures, in different levels, will be recurrent and ongoing for a long time. Due to this fact, compensatory social aid policies should also be developed in order to reduce the effects of a possibly long-lasting economic turn-down. Recent papers have discussed this matter \cite{ahamed2020role, zacchi:hal-02881690}.

\section*{Acknowledgments}
The Authors acknowledge the financial support of National Council for Scientific and Technological Development (CNPq, Brazil) under grants $304032/2019-0$ and $201143/2019-4$ (PhD Program Abroad). The Authors also thank Saulo B. Bastos and Daniel O. Cajueiro for previous collaborations and discussions.

 \subsection*{Notes}
The authors report no financial disclosure nor any potential conflict of interests.
 
\bibliographystyle{model5-names}
\bibliography{isat_covid}

\end{document}